\documentclass[usenatbib]{mnras}

\usepackage{tabularx}

\usepackage{pdflscape} %
\usepackage{graphicx}
\usepackage[space]{grffile}
\usepackage{latexsym}
\usepackage{amsfonts,amsmath,amssymb}
\usepackage{url}
\usepackage[utf8]{inputenc}
\usepackage{hyperref}
\hypersetup{colorlinks=false,pdfborder={0 0 0}}
\usepackage{textcomp}
\usepackage{longtable}
\newcommand{\LDWARFSB}{543}
\newcommand{\LDWARFSF}{1010}
\newcommand{\TDWARFSB}{10}
\newcommand{\TDWARFSF}{58}
\newcommand{\NTOT}{1885} 
\newcommand{\NGLM}{1317} 
\newcommand{\NFND}{328} 
\newcommand{\NREM}{3} 
\newcommand{\NCAT}{321} 
\newcommand{\NPMF}{145} 
\newcommand{\NDUP}{8} 
\newcommand{\NGDF}{7} 
\newcommand{\NIND}{6} 
\newcommand{\NGTL}{10} 
\newcommand{\PERCENT}{45}
\newcommand{\NGJF}{304} 
\newcommand{\NHRD}{49}
 
\newcommand{\TOTCPMS}{32}
\newcommand{\OLDCPMS}{17}
\newcommand{\NEWCPMS}{15}

\newcommand{\G}{\textit{Gaia}}

\usepackage{natbib}

\title[Known LT dwarfs and the \G~DR1]{The  \G~Ultracool Dwarf Sample. I. Known L and T dwarfs and the first  \G~data release}

\author[R. L. Smart et al.]
{R. L. Smart$^{1,2}$\thanks{\tt E-mail: smart@oato.inaf.it, Leverhulme  Visiting Professor},
F. Marocco$^2$,
J.~A.~Caballero$^{3,4}$,
H. R. A. Jones$^2$,
D.~Barrado$^{4}$,
\newauthor
J. C. Beam\'in$^{5,6}$,
D. J. Pinfield$^2$,
L.~M.~Sarro$^{7}$
\\
$^{1}$Istituto Nazionale di Astrofisica, Osservatorio Astrofisico di Torino, Strada Osservatorio 20, 10025 Pino Torinese, Italy\\
$^{2}$School of Physics, Astronomy and Mathematics, University of Hertfordshire, College Lane, Hatfield AL10 9AB, UK\\
$^{3}$Landessternwarte K\"onigstuhl, Zentrum f\"ur Astronomie der Universit\"at Heidelberg, K\"onigstuhl 17, D-69117 Heidelberg, Germany\\
$^{4}$Centro de Astrobiolog\'ia, Dept. Astrofisica INTA-CSIC, ESAC campus, Camino Bajo del Castillo s/n, E-28692, Villanueva de la Ca\~nada, Madrid, Spain\\
$^{5}$Instituto de F\'isica y Astronom\'ia, Universidad de Valpara\'iso, Ave. Gran Breta\~na, 1111, Valpara\'iso, Chile\\
$^{6}$Millennium Institute of Astrophysics, Santiago, Chile\\
$^{7}$Departamento de Inteligencia Artificial, ETSI Inform\'atica, UNED, Juan del Rosal, 16 28040 Madrid, Spain
}

\begin{document}



\maketitle

\begin{abstract}
{We identify and investigate known ultracool stars and brown dwarfs that are
  being observed or indirectly constrained by the \G~mission. These objects will
  be the core of the \G~ultracool dwarf sample composed of all dwarfs later
  than M7 that \G~will provide direct or indirect information on.  We match
  known L and T dwarfs to the \G~first data release, the Two Micron All Sky
  Survey and the {\it Wide-field Infrared Survey Explorer} AllWISE survey and
  examine the \G~and infrared colours, along with proper motions, to improve
  spectral typing, identify outliers and find mismatches.  There are {\NCAT} L
  and T dwarfs observed directly in the \G~first data release, of which
  {\NGTL} are later than L7.  This represents \PERCENT\,\% of all the known LT
  dwarfs with estimated \G~$G$ magnitudes brighter than 20.3\,mag.  We
  determine proper motions for the {\NCAT} objects from \G~and the Two Micron
  All Sky Survey positions. Combining the \G~and infrared magnitudes provides
  useful diagnostic diagrams for the determination of L and T dwarf physical
  parameters. We then search the Tycho-\G~astrometric solution \G~first data
  release subset to find any objects with common proper motions to known L
  and T dwarfs and a high probability of being related. We find \NEWCPMS~new
  candidate common proper motion systems.}
\end{abstract}

\section{Introduction}
\label{section.introduction}

\begin{figure}
\centering
\includegraphics[width=0.49\textwidth]{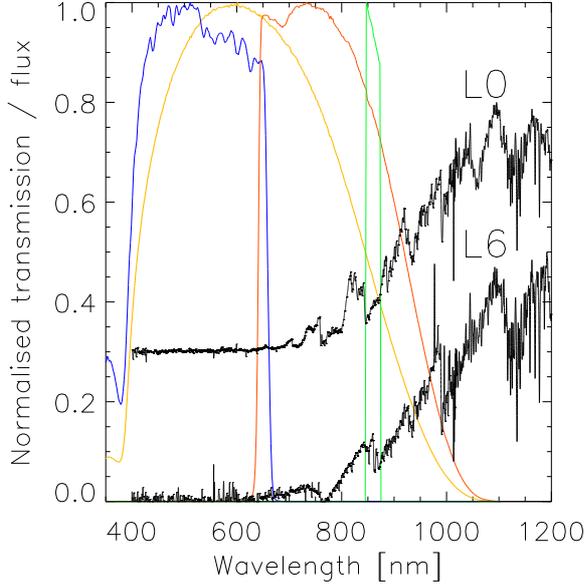}
\caption{Normalised instrumental transmission for \G~$G$ (yellow), $G_{\rm
    BP}$ (blue), $G_{\rm RP}$ (red) and $G_{RVS}$ (green ) filters and
  optics. The spectra marked L0 and L6 are from X-Shooter 
  for the L0 dwarf 2MASS J23440624--0733282 and the L6 dwarf 2MASS
  J00065794--6436542, arbitrarily scaled and off set from each other.  }
\label{passbands}
\end{figure}

\G~is observing over a billion objects in our Galaxy and is revolutionizing
Astronomy in many areas \citep{2016A&A...595A...1G}.  One of these areas is
the study of the bottom of the main sequence and beyond.  L and T (hereafter
LT) dwarfs are very cool faint objects that are either substellar or at the
stellar-substellar boundary \citep{1997A&A...327L..25D, 1999AJ....118.1005M,
  1999ApJ...519..802K, 2006ApJ...637.1067B, 2014AJ....147...94D}.  In the
billion-object catalogue of \G, there will be direct observations of about a
thousand LT dwarfs \citep{2013A&A...550A..44S, 2014MmSAI..85..649S}.  This
sample, even though it is relatively small, will be more homogeneous,
accurate, complete and larger than the current catalogue of known L and early T
dwarfs with measured parallactic distances.

\begin{table}
 \caption[]{Distance limits for L0 to T7 spectral types using
  Eq.~\ref{newgvsspt}. D$_{G<20.3} = $distance limit assumming $G<$20.3
  mags and D$_{G<20.7}$ for $G<$20.7 mags. }
\label{table:distancelimits}
\begin{tabular}{|lcc| lcc|}
   \hline
L  & D$_{G<20.3}$ & D$_{G<20.7}$ &   T  & D$_{G<20.3}$ & D$_{G<20.7}$ \\
SpT     &   (pc)     &   (pc) &  SpT     &   (pc)     &   (pc)           \\
            \hline
L0 & 69 & 82 & T0 & 12 & 14 \\ 
L1 & 55 & 67 & T1 & 12 & 14 \\ 
L2 & 45 & 54 & T2 & 12 & 14 \\ 
L3 & 36 & 44 & T3 & 12 & 14 \\ 
L4 & 29 & 35 & T4 & 11 & 14 \\ 
L5 & 24 & 29 & T5 & 10 & 12 \\ 
L6 & 19 & 23 & T6 & 8 & 10  \\ 
L7 & 16 & 19 & T7 & 6 & 7   \\ 
L8 & 13 & 15 & T8 & 3 & 4   \\ 
L9 & 10 & 12 & T9 & 2 & 2   \\ 
 \hline
 \end{tabular}       
\end{table}

The reason for the relative paucity of LT dwarfs in the \G~observations is
because they emit predominantly in the infrared and are very faint in the
\G~bands (see Fig. \ref{passbands}).
However, \G~will provide a magnitude-limited complete sample of the early LT
spectral types in the solar neighbourhood. The nominal \G~$G$ magnitude limit
is 20.7\,mag and we expect the mission to be complete to $G=20.3$ mag
\citep{2016A&A...595A...2G}. Internal validation, with models and clusters,
finds a completeness of 50\% at $G=20.3$\,mag for the \G~first data release
(hereafter DR1)\footnote{\url{http://gaia.esac.esa.int/documentation/GDR1/}}.
In Table \ref{table:distancelimits} we report the distance limits for L0 to T9
objects with a $G$ limit of 20.3\,mag and 20.7\,mag using Eq. \ref{newgvsspt}
developed in Section \ref{section:identifying}.  In addition to
solar-metallicity LT dwarfs, \G~will also provide a volume limited sample of
old thick disk or halo L-type sub-dwarfs, and young LT objects in the solar
neighbourhood.

Other LT and even cooler Y dwarfs \citep{2011ApJ...743...50C} will be
indirectly detected in \G~observations, for example, as { low-mass companions
  in} unresolved binary systems \citep{ 2013ApJ...767..110P,
  2014MNRAS.437..497S, 2014MNRAS.445.2106L, 2015AJ....149..104B,
  2016AJ....152..112M}, and as gravitational microlenses \citep{
  2002MNRAS.331..649B, 2011A&A...536A..50P, 2014ApJ...782...89S,
  2014sf2a.conf..269R}.  \G~will constrain other LT and Y dwarfs in common
proper motion (CPM) systems of wide binaries or moving groups where distances
and kinematics of the brighter members, visible to \G, can be matched to the
fainter objects with kinematics found from other surveys.

Ultracool dwarfs (UCDs) are defined as objects later than M7
\citep[see][]{2001udns.conf.....J}.  We have begun a systematic project to
catalogue and characterise the cooler part of the \G~Ultracool Dwarf sample
(hereafter GUCDS), being all L, T and Y dwarfs that \G~will directly observe or
indirectly constrain. The GUCDS will be the primary sample in the near future
to test atmospheric models and evolution scenarios, and to derive fundamental
properties of objects at the end of the main sequence.

Here we find the LT dwarfs directly observed by \G~as isolated
objects with an identifiable entry in the \G~DR1 and we find those LT dwarfs
in CPM systems with the \G~DR1 subset with astrometric solutions. In
Section~\ref{section:catalogue} we describe the L, T and Y dwarf input catalogue
used to search the \G~DR1; in Section~\ref{section:identifying3} we describe
the production of the GUCDS catalogue of known matched LT dwarfs; in
Section~\ref{section:cpm} we describe
the discovery of new CPM candidates;  in  Section~\ref{section:results}
we discuss the two catalogues in various magnitude, colour, and proper
motion parameter spaces and in the last section we summarise the results.


\section{Catalogue of known L, T and Y dwarfs}
\label{section:catalogue}


\subsection{Input catalogue }

   \begin{table*}
      \caption[]{The GUCDS input catalogue$^a$.} 
         \label{table:input}
     $$ 
         \begin{tabular}{ll cc l cc c}
            \hline
            \hline
            \noalign{\smallskip}
GUCDS          		& Name                         			&  $\alpha$	& $\delta$ 	& Spectral			& $\mu_{\alpha} \cos{\delta}$	& $\mu_{\delta}$	& $J$ \\
ID           		&                        					& [deg]             	& [deg]		& type       			& [mas/yr] 					& [mas/yr]          		& [mag] \\
            \noalign{\smallskip}
            \hline
            \noalign{\smallskip}
J0000+2554 	& 2MASS J00001354+2554180	& 0.0564165	& +25.905000 	& T4.5$^2$ 		& +6$\pm$19   				& +130$\pm$22 	& 15.063\\%
J0001+1535 	& 2MASS J00011217+1535355 	& 0.3007080	& +15.593194 	& L4$^2$ 			& +150$\pm$19  			& --169$\pm$19 	& 15.522\\%
J0001--0841 	& 2MASS J00013166-0841234 	& 0.3830416	& --8.690806	& L1 p(blue)$^2$ 	& +331$\pm$14  			& --299$\pm$14 	& 15.712\\%
J0002+2454 	& 2MASS J00025097+2454141 	& 0.7123755	& +24.903917	& L5.5$^2$ 		& +2$\pm$23   				& --36$\pm$29 		& 17.165\\%
J0004-6410 	& 2MASS J00040288-6410358 	& 1.0120000	& --64.176611 	& L1 $\gamma$$^1$ & +64$\pm$5 				& --47$\pm$12 	 	& 15.786\\%
...\\
         \noalign{\smallskip}
            \hline
         \end{tabular}
     $$ 
$^a$ The full table of 1886 entries is available online at TO BE FILLED IN BY
         MNRAS, references for each variable are in the online
         version. Superscripts $^1$ and $^2$ indicate that the adopted
         spectral type were measured in the optical and near infrared,
         respectively.
\end{table*}

LT dwarfs seen by \G~will all be nearby ($d <$ 82\,pc;
Table~\ref{table:distancelimits}) and, therefore, have significant proper
motions. With this in mind, we used as the starting point for our input
catalogue of known late M, L, T and Y dwarfs the online census being kept by
J.~Gagn\'e\footnote{\url{https://jgagneastro.wordpress.com/list-of-ultracool-dwarfs/}}.
This included objects from the
Dwarfarchives\footnote{\url{http://www.dwarfarchives.org/}}, the work of
\cite{2012ApJS..201...19D}, and the PhD thesis catalogue of
\cite{2014yCat.5144....0M}. To this compilation, we added the objects in
\cite{2015MNRAS.449.3651M} and \cite{2016ApJS..225...10F}.  {We did not
  include} the significant number of UCD candidates with
photometry-based spectral types \citep[e.g. ][]{2012MNRAS.427.3280F,
  2014MNRAS.443.2327S, 2016A&A...589A..49S}, since they are mostly too faint
for \G~and do not yet have proper motion estimates.

We confine our sample to all objects that have an optical or infrared spectral
type equal to or later than L0 or are young late type M dwarfs that are probable
brown dwarfs (e.g. TWA 27 A \cite{2002ApJ...575..484G}).
These objects cover a large age range and include objects in the stellar,
brown dwarf and giant-planet regimes. While \G~is only observing directly a
few objects later than L7, we included all published L, T and Y dwarfs, as the
same list is used to search for common proper-motion objects in the
\G~DR1. Most UCDs (in particular late-M and early L
dwarfs) have been classified using both their optical and near-infrared
spectra, leading to two different and sometimes discordant spectral
types. When we had to choose a spectral type, for example to calculate
spectroscopic distances, when available we adopted optical spectral 
types for late-M and L
dwarfs, since the wealth of spectral lines and bands in the
5\,000--10\,000\,{\AA} wavelength range makes the classification more
accurate, while for T dwarfs we use their near-infrared spectral type
following similar considerations.

The current version of the input catalogue contains \NTOT~entries.
In Table~\ref{table:input} we list the short name, discovery name,
equatorial coordinates, adopted spectral type, proper motions and
$J$-band magnitude of the first five UCDs of the list. The
full GUCDS input catalogue with references for each variable is
available online at {\bf TO BE FILLED BY MNRAS. } This list will evolve, and be
updated and maintained, as part of the GUCDS initiative in the MAIA database
\citep{2014MmSAI..85..757C}.


\subsection{Predicted $G$ magnitude}

To first estimate a \G~$G$ magnitude for the input catalogue, we used the
procedure developed in \citet{2014MmSAI..85..649S}.  Briefly, we combined the
Two Micron All Sky Survey \citep[hereafter 2MASS; ][]{2006AJ....131.1163S} $J$
magnitudes, Sloan Digital Sky Survey \citep[hereafter SDSS;
][]{2000AJ....120.1579Y} colours as a function of spectral type from Table 3
in \citet{2002AJ....123.3409H}, and colour transformations between
\G~photometry and the SDSS system from \citet{LL:CJ-041} to find a predicted
$G$ magnitude. To this table we fitted a simple linear polynomial of 
predicted $G$ magnitudes as a function of spectral type and $J$ magnitude
to obtain:
\begin{equation}
G_{\rm pred} = J - 12.63 + 0.244~{\rm SpT},
\label{originalGvssp}
\end{equation}
where SpT is the numerical representation of the LT types from 70 to 89
equivalent  to L0 to T9.

The Jordi (2012) \G-to-SDSS transformations were based on main sequence stars
in the colour range $g-r$ = (--0.5, 7.0)\,mag.  There will be a systematic
error in Eq.~\ref{originalGvssp} due to the difference between M and LT dwarf
spectral energy distributions, but we estimated this to be less than 0.2\,mag
by extrapolating the difference between M giants and dwarfs in the
transformation construction.  The transformation is imprecise because of the
multiple steps, the use of 2MASS magnitudes and this systematic
error. However, Eq.~\ref{originalGvssp} was only used to constrain the objects
that we search for, so we considered it sufficient.  From the input list we
searched {the} \G~DR1 for all objects with a predicted magnitude $G_{\rm
  pred}<23$\,mag. Since the nominal DR1 limit is $G=20.7$\,mag, this allowed
for significant random or systematic errors {in our relationship and its
  parameters ($J$, SpT)}.  Of the original \NTOT~objects, \NGLM~were brighter
than this conservative $G_{\rm pred} <$ 23\,mag cut. In the final matched
catalog the faintest object had a $G_{\rm pred}=22.6$\,mag.


\section{Identification of DR1 matches}
\label{section:identifying3}

\subsection{Initial matching}
\label{section:identifying}

\begin{figure}
\centering
\includegraphics[width=0.49\textwidth]{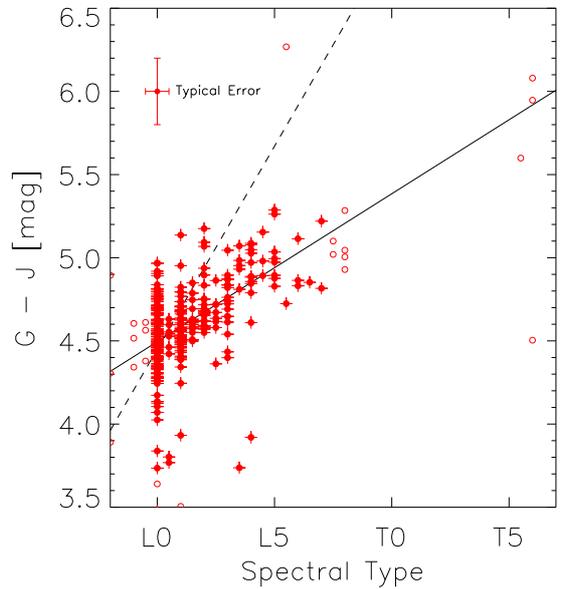}
\caption{All matched \NFND~objects $G-J$ magnitude vs spectral type plotted as
  open circles.  The \NGJF~filled circles are L0--7 dwarfs with one matched DR1
  entry within 3\,arcsec and a zero DR1 {\em duplicated\_source} flag used to
  find the solid line, Eq.~\ref{newgvsspt}.  The dashed line is the $G_{\rm
    pred}$ found from Eq.~\ref{originalGvssp}. See Section
  \ref{section:identifying} for details. Typical errors bars are shown.}
\label{sptvsGmJ1}
\end{figure}

Since our input objects generally have high proper motions, and both the
ground-based GUCDS input catalogue and the DR1 are of different epochs and
with varying completeness, the identification of the LT dwarfs in the DR1
required a careful cross match.  For each object, we matched the published
position moved to the the DR1 epoch using the proper motions in our input
list.  We found that \NFND~of the \NGLM~UCDs had a DR1 entry
within a matching radius of 3\,arcsec. We also considered other matching radii
both smaller (2\,arcsec) and larger (5\,arcsec), and found 3\,arcsec to be the
best compromise between too many false matches and missing true high proper
motion objects.  
Of these, \NIND~objects had more than one DR1 match within
3\,arcsec, and \NDUP~had a non-zero {\em duplicated\_source} flag in the DR1,
which indicates that during the \G~processing the source at some point was
duplicated.

We then determined a new relationship for estimating $G$ magnitudes from 2MASS
$J$ magnitudes and tabulated spectral types. We selected the \NGJF~
cross-matched L dwarfs that ($i$) had only one DR1 match within 3\,arcsec,
($ii$) had a zero \G~{\em duplicated\_source} flag, and ($iii$) were earlier
than L7.  For this sub-sample, using least squared absolute deviation we found
the first-order polynomial relationship between the colour $G-J$ and spectral
type as:
\begin{equation}
    G_{\rm est}= J - 1.098  + 0.080~{\rm SpT}
\label{newgvsspt}
\end{equation}
valid for SpT = 70 to 77, i.e. L0 to L7.

The colour-spectral type diagram in Fig.~\ref{sptvsGmJ1} illustrates the
measured $G$  minus $J$ magnitudes with lines that represent $G_{\rm pred}$
(Eq.~\ref{originalGvssp}), and this new robust fit, $G_{\rm est}$
(Eq.~\ref{newgvsspt}).  The new relation in Eq.~\ref{newgvsspt} is much
flatter than Eq.~\ref{originalGvssp}. We found  \NGDF~objects in that have a 
measured and estimated $G$ difference, $\Delta G = |G-G_{\rm est}|$, larger 
than 1\,mag.  While the
underestimation of the $G-J$ for the T6 object indicates extrapolating the fit
beyond L7 provides uncertain results, we only used this $\Delta G$ flag as an
indicator of possible problems.

\begin{figure}
\centering
\includegraphics[width=0.49\textwidth]{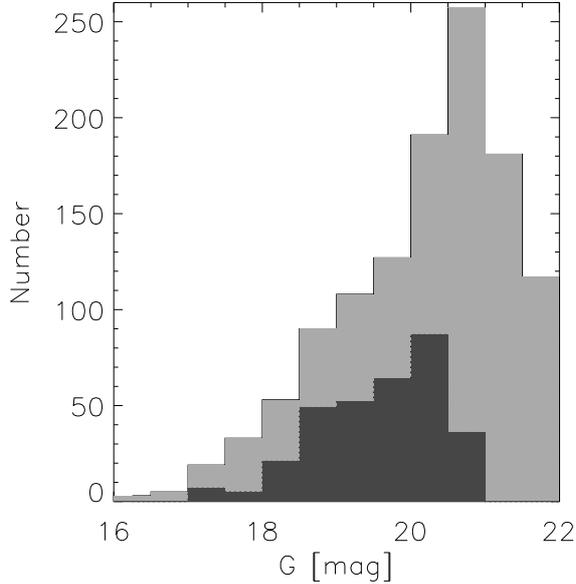}
\caption{Distribution histograms of $G_{\rm est}$ for the input catalogue (light
  grey) and of \G~measured $G$ magnitudes (dark grey).}
\label{sptvsGmJ}
\end{figure}
%

\begin{figure}
\centering
\includegraphics[bb=100pt 1pt 768pt 300pt,width=0.60\textwidth]{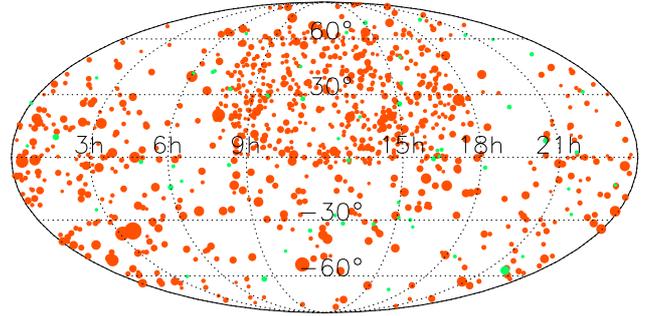}
\caption{The equatorial distribution of ~\LDWARFSF~ L (red) and ~\TDWARFSF~ T (green)
  dwarfs with $G_{\rm est}<21.5$\,mag.  The size of the symbol indicates the
  $G_{\rm est}$ magnitude -- larger is brighter.
}
\label{skydist}
\end{figure}

Fig. \ref{sptvsGmJ} is the distribution of the input catalogue in $G_{\rm
  est}$ magnitudes using Eq.~\ref{newgvsspt} for the input catalogue and the
\NFND~matched objects with measured $G$ magnitudes. The degree of completeness
varies greatly from 0\,\% in the bright bins below $G=16.5$\,mag and the faint
bins beyond $G=21$\,mag, to over 50\,\% at $G=19$\,mag. The brightest bins have the
objects with the highest proper motions so are systematically effected by 
\G~observation matching problems \citep[][]{2016A&A...595A...3F}.  In general,
the incompleteness can be attributed to objects that were excluded from DR1,
matching problems due to imprecise positions and/or proper motions or
mis-classifications in the GUCDS input catalogue leading to over estimated
$G_{\rm est}$ magnitudes.

In total, there are \LDWARFSF~L and \TDWARFSF~T dwarfs brighter than $G_{\rm
  est}=21.5$\,mag, and \LDWARFSB~L and \TDWARFSB~T dwarfs brighter than
$G=20.3$\,mag. In contrast, Smart (2014) predicted only two T dwarfs to
$G=20.3$\,mag. The higher number estimated in this work is due to the
systematic underestimation of the $G_{\rm pred}$ (Eq.~\ref{originalGvssp})
used in Smart (2014) with respect to the $G_{\rm est}$ (Eq.~\ref{newgvsspt}). 
Using a more theoretical approach, Sarro et al (2013) predicted of the 
order of 10 T dwarfs brighter than $G=20.0$\,mag.

In Fig. \ref{skydist} we plot the sky distribution of all the input
catalogue with $G_{\rm est} <$ 21.5\,mag using Eq. \ref{newgvsspt}.  
The region of over density in the northern hemisphere is
from the SDSS footprint, and is probably representative of a complete sky
\citep{2010AJ....139.1808S}. However, the Galactic plane is incomplete, as most of
the LT dwarfs discovered to date have been via photometric selection, and the
crowding in the plane makes this difficult.

Of the \NIND~objects that had more than one DR1 entry within
3\,arcsec some may be due to binarity or a background object near to
the catalogue dwarf, but most are due to multiple entries in the DR1 \cite[see
  Section~4 in][]{2016A&A...595A...2G}.  It is estimated that the multiple entries in the
DR1 catalogue is a few percent \citep{2016A&A...595A...3F}, consistent with this finding.

In the GUCDS input catalogue, objects either have published proper motions or
we estimated them from the 2MASS and {\it Wide-field Infrared Survey Explorer}
AllWISE\footnote{\url{http://irsa.ipac.caltech.edu/data/download/wise-allwise/}}
positions \citep{2010AJ....140.1868W}. We compared these input values with a
derived proper motion from the difference of the \G~DR1 and the 2MASS
position.  When the magnitude of the proper motions differed by more than
20\,\%, we flagged the object. This resulted in \NPMF~objects being flagged,
i.e. $\sim$ 50\,\%. This high percentage is not unexpected given that both proper
motions are of low precision and the parallactic motion of the object is
unknown.


\subsection{Identifying mismatches}

\begin{figure}
\centering
\includegraphics[width=0.41\textwidth]{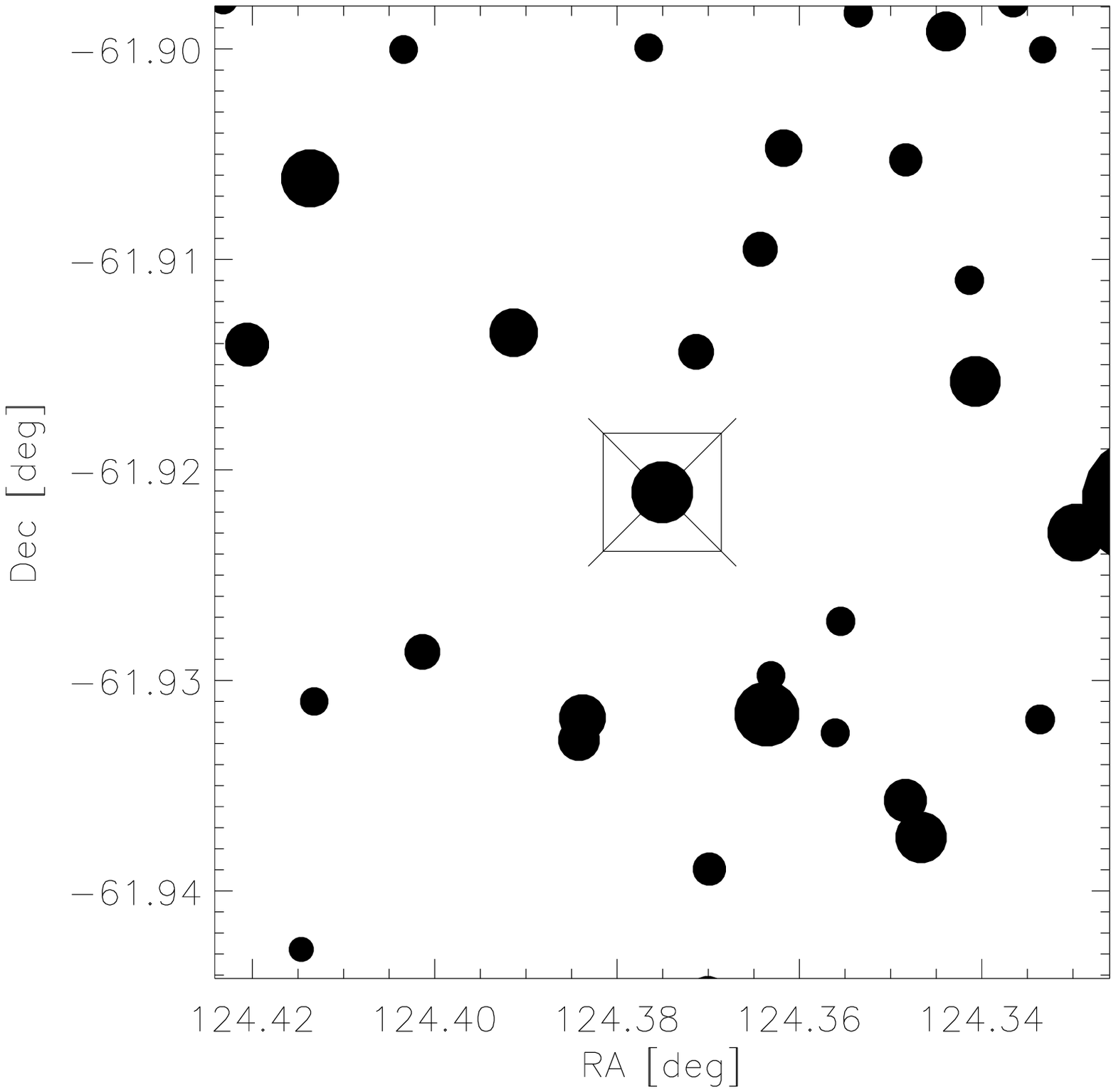}
\includegraphics[width=0.41\textwidth]{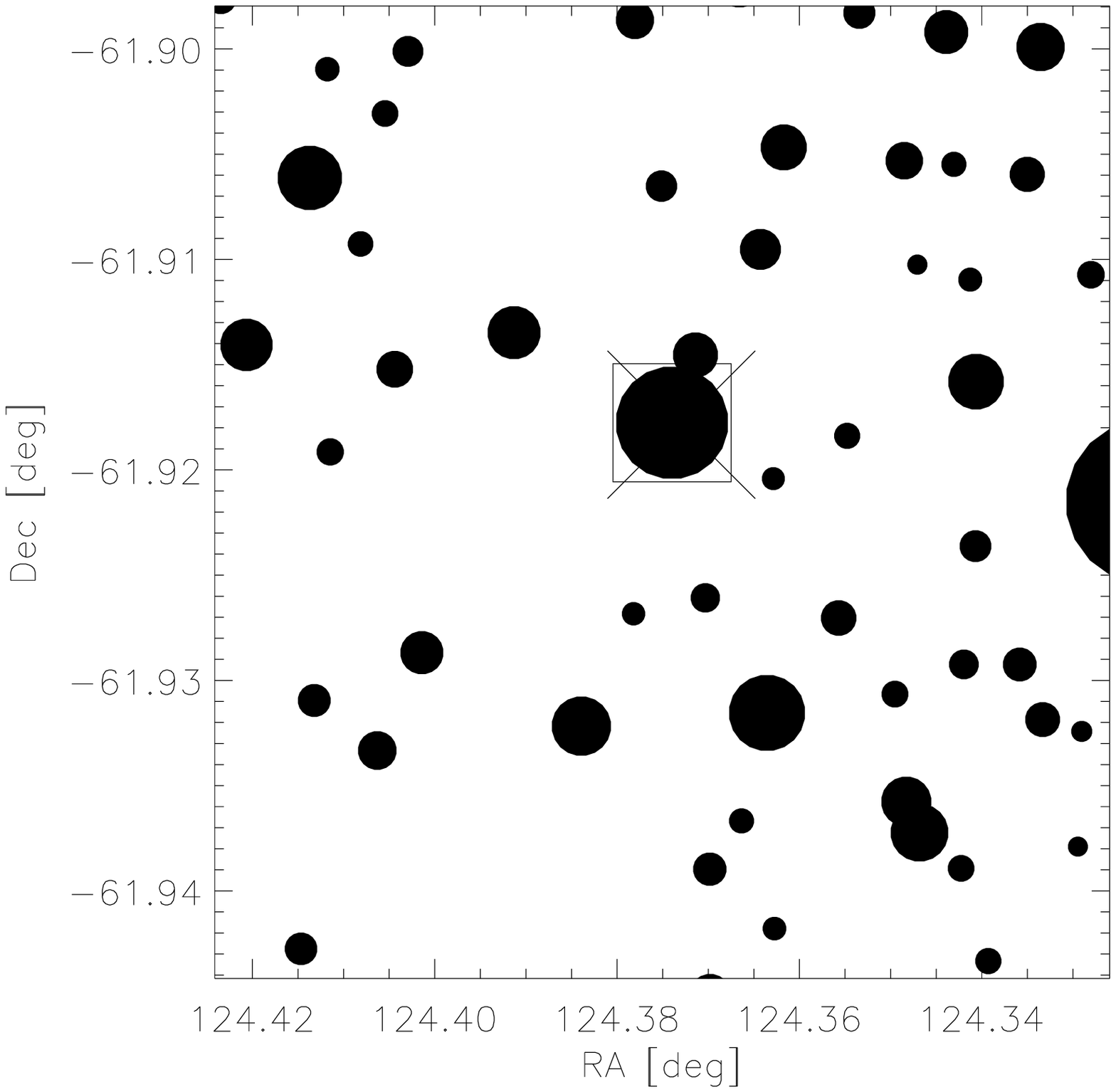}
\includegraphics[width=0.41\textwidth]{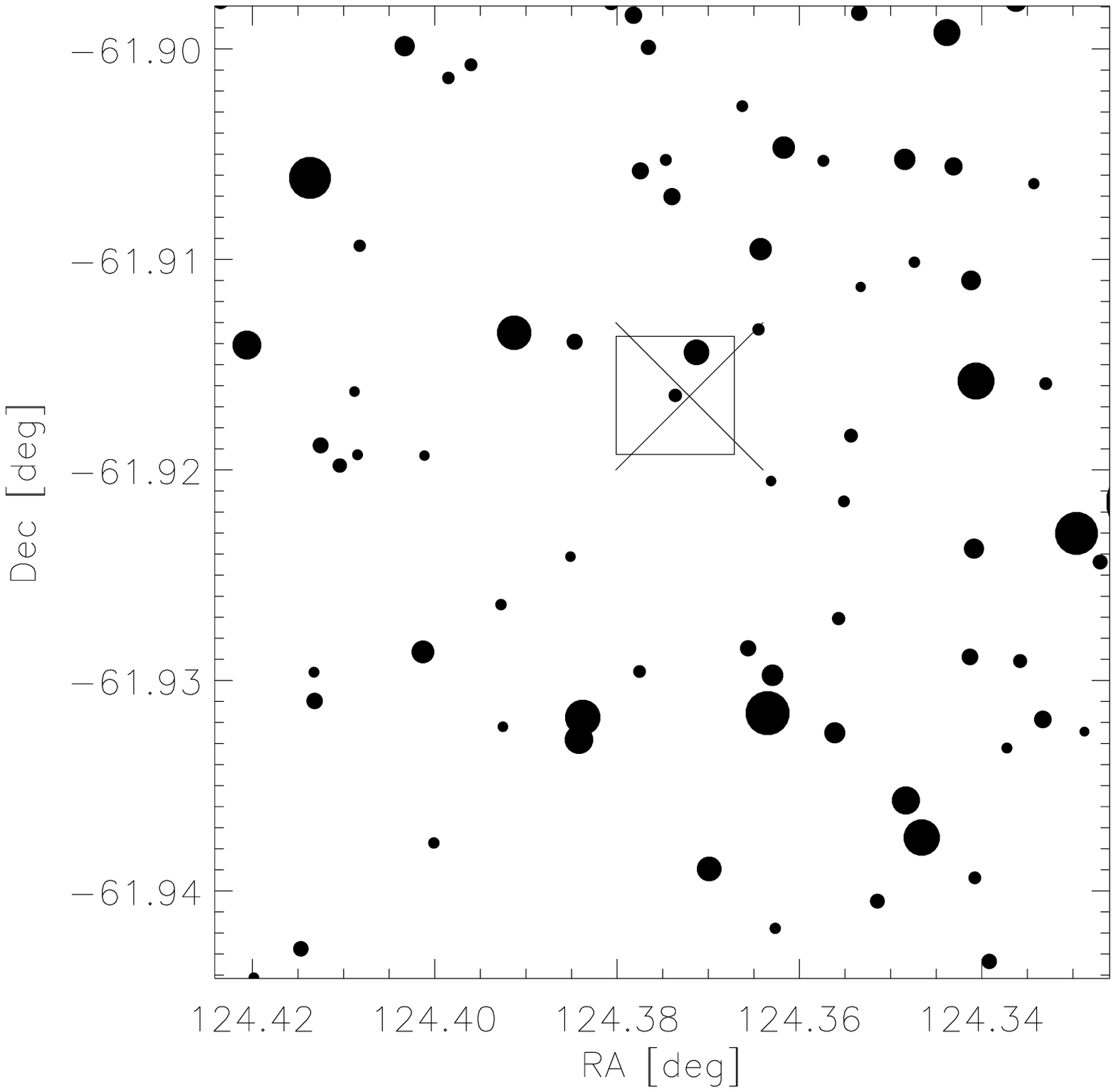}
\caption{Sky plots of a 2.5$\times$2.5\,arcmin field around J0817-6155 in
  2MASS $J$ (top panel, epoch 2000.0), AllWISE $W2$ (middle, epoch 2010.6) and
  DR1 $G$ (bottom, epoch 2015.0). Brighter objects are plotted as larger
  symbols. The ``X'' matches the predicted position based on the literature
  position and proper motions, the 20\,arcsec square is centred on the
  respective catalogue source.}
\label{J0817-6155}
\end{figure}

Since \G~does not produce images (in general), we cannot perform the usual visual
confirmation to look for mismatches. We confined our examination for
mismatches to the catalogue maps and various flags.  For each target
we constructed a quality assurance output including: the flags for
proper motion and $|G-G_{est}|$ magnitude differences; number of DR1
observations; the 2MASS images; positions and magnitudes from the
2MASS, AllWISE and DR1 catalogues; the input spectral types;
parallaxes when published; input comments (e.g. known binarity or
subdwarf); literature and calculated proper motions; and plots of the
fitted proper motions and sky maps for the field in both 2MASS,
AllWISE and DR1 catalogues. 

In Fig. \ref{J0817-6155} we show an example of the sky distribution plots for
the field around the T6 J0817-6155 \citep{2010APJ...718L..38A}.  The slight
misalignment between the cross and the square in the DR1 panel is due to
imprecise starting proper motions from the input catalogue.  When needed, we
also examined online ground based images of the fields.

We examined all \NFND~targets to see if any of the candidate were obvious
mismatches. In particular, we paid special attention to the objects with large
magnitude differences, multiple DR1 entries within 3\,arcsec, and the
\NGTL~objects later than L7. Of the \NFND~targets we identified \NREM~objects
that we believe are mismatches and are listed in Table \ref{mismatches}. Most of these
had $\Delta G>1.0$ mag, proper motion differences larger than 20\,\%, and/or
visual inspection of the field did not allow an unambiguous identification.
The \G~second data release is expected to resolve these ambiguities.

\begin{table*}
   \caption[]{Input catalogue objects with mismatches in   \G~DR1.} 
      \label{mismatches}
      \begin{tabular}{lcll}
         \hline
         \hline
         \noalign{\smallskip}
  Name         &    \G~DR1 Source ID & $\Delta G $[mag] & Remarks \\
         \noalign{\smallskip}
         \hline
         \noalign{\smallskip}
  WISEPA J062720.07-111428.8$^1$ &  3000505938722626560 & -1.4 & Probable satellite of galaxy USNO-B1 0787-0079432	       \\
    USco J160714.79-232101.2$^2$ &  6242316978518793856 & -1.4 & Star or galaxy USNO-B1 0666-0359684			       \\
    USco J163919.15-253409.9$^3$ &  6046485475060342528 & -1.1 & Star or galaxy USNO-B1 0644-0382256			       \\
    WISE J170745.85-174452.5$^4$ &  4135505777467362304 & -2.6 & In crowded region towards Galactic centre \\
      2MASS 18000116-1559235$^5$ &  4145132555890843520 &  1.3 & In crowded region of low Galactic latitude \\
    WISE J200403.17-263751.7$^6$ &  6850032688873127424 & -0.9 & Star or galaxy USNO-B1 0633-0938809			       \\
    WISE J223617.59+510551.9$^4$ &  1988335902592180608 & -0.3 & Star or galaxy USNO-B1 1410-0455016                         \\
      \noalign{\smallskip}
         \hline
      \end{tabular}
Discovery references:   $^1$\cite{2011APJS..197...19K},
  $^2$\cite{2007MNRAS.379.1423L},
  $^3$\cite{2009A&A...505.1115L},
  $^4$\cite{2013APJ...777...36M},
  $^5$\cite{2012MNRAS.427.3280F},
  $^6$\cite{Thompson2013}
\end{table*}


\subsection{DR1 multiple matches}

There are \NIND~LT dwarfs with multiple matches within 3\,arcsec. Three,
J0257-3105 \citep{2008APJ...689.1295K}, J0543+6422 \citep{2008AJ....136.1290R}
and J1515+4847 \citep{2003IAUS..211..197W}, are matched to DR1 entries within
only 1\,arcsec and the DR1 entries have very similar magnitudes. The matches
to these three are probably duplicated DR1 entries \cite[see discussion
  in][]{2016A&A...595A...3F}, and we adopted the DR1 entry with the highest
number of observations. The candidate J1203+0015 \citep{2000AJ....119..928F}
is matched to two entries with significantly different magnitudes (both
fainter than the estimated \G~magnitude) so it is probably either close to a
background object, or \G~has resolved the dwarf into a binary system with a
0.3\,arcsec separation.  The targets J1606-2219 and J1607-2211
\citep{2007MNRAS.379.1423L} have fainter detections 2-3\,arcsec away, which we
believe to be background objects. In these last three cases we adopted the
match closest to the predicted LT dwarf position.

\subsection{Completeness}
If we consider input catalogue  objects with $G_{\rm est} < 20.3$\,mag we find only
\PERCENT\,\% in the \G~DR1. This incompleteness is due primarily to the quality assurance
cuts of \G~which are: $ N > 5, \epsilon_i < 20$\,mas, and $\sigma_{\rm pos,max} <
100$\,mas, where $N$ is the number of field-of-view transits used in the
solution, $\epsilon_i$ is the excess source noise, and $\sigma_{\rm pos,max}$
is the semi-major axis of the error ellipse in position at the reference epoch
\citep[from Section 5 in][]{2016A&A...595A...4L}. In addition, we required all 
included objects to have valid photometry. The number of field-of-view transits led to
a systematic incompleteness that follows the scanning law and can be seen in
the sky plots of \G~
DR1\footnote{\url{http://sci.esa.int/gaia/58209-gaia-s-first-sky-map}}.
Importantly for these objects, the cyclic processing does not yet use internal
proper motions to update the position of objects during the matching, so the
correct matching of high proper motion objects is deficient
\citep[][]{2016A&A...595A...3F}. Given the documented incompleteness of 50\,\% at
$G=20.3$\footnote{\url{http://gaia.esac.esa.int/documentation/GDR1/}}\,mag,
and the very high proper motion of most bright LT dwarfs, we consider the
success rate of \PERCENT\,\% to be reasonable. The matching for DR2 will
include internal proper motions, so it will not have this deficiency.

\subsection{\G~observed L and T dwarfs Catalogue}

\begin{table*}
   \caption[]{New parameters for the GUCDS-DR1 catalogue.$^a$} 
      \label{catalog} 
      \begin{tabular}{lccccccc}
         \hline
         \hline
         \noalign{\smallskip}
Short    &   \G~              &   $\alpha, \delta$          &
$\mu_{\alpha}$cos$\delta $, $\mu_{\delta}$ & $G $ & $\Delta G$ & N$_{obs}$, N$_3$, F$_\mu$ \\
 Name    & Source ID           & [deg]       &    [mas/yr]      &      [mag] &         [mag] &       \\
         \noalign{\smallskip}
         \hline
         \noalign{\smallskip}
 J0006-1720 & 2414607592787544320 &    1.585253567, -17.347415311 &   -41$\pm$ 11,    0$\pm$ 12 &  20.525$\pm$0.043 &   0.163 &  97,0,1\\
 J0006-0852 & 2429054454021227648 &    1.704590901,  -8.880825977 &   -61$\pm$ 12, -324$\pm$ 12 &  18.485$\pm$0.008 &  -0.078 & 239,0,0\\
 J0006-6436 & 4900323420040865792 &    1.742217614, -64.615322431 &    82$\pm$  5,  -65$\pm$ 13 &  17.988$\pm$0.011 &   0.103 & 145,0,0\\
 J0016-1039 & 2428008410441149824 &    4.156258155, -10.653818836 &  -111$\pm$ 12, -193$\pm$ 12 &  20.028$\pm$0.022 &   0.073 & 185,0,0\\
 J0018-6356 & 4900453540369793408 &    4.695160455, -63.938248147 &   324$\pm$  6, -381$\pm$ 13 &  19.774$\pm$0.025 &  -0.110 & 108,0,0\\
 ... \\
         \noalign{\smallskip}
            \hline
         \end{tabular}
     $^a$ {Equatorial coordinates and apparent magnitudes are from \G~DR1 at
        epoch J2015.0, while proper motions were computed by us after using
        2MASS and \G~astrometry. $ \Delta G = G - G_{\rm est}$; N$_{obs}$ =
        number of \G~observations; N$_3$ = number of DR1 entries within
        3\,arcsec and F$_\mu$ is a flag to indicate if the calculated proper
        motion was within 20\% of the input value. The full table of 321 LT
        dwarfs with other supporting magnitudes and references are available
        online at TO BE FILLED BY MNRAS.}
   \end{table*}

\begin{figure}
\includegraphics[bb=100pt 1pt 768pt 300pt,width=0.60\textwidth]{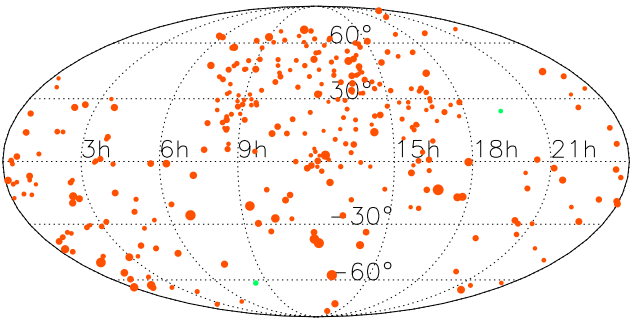}
\caption{Same as Fig. \ref{skydist}, but for the  \NCAT~L (red) and T (green)
  dwarfs with an entry in \G~DR1. }
\label{skydistfound}
\end{figure}

We produced a catalogue of the parameters for the  \NCAT~ L and T dwarfs with
a reliable entry in \G~DR1,
which are distributed as shown in Fig. \ref{skydistfound}. This will
be actively updated on-line along with the input catalogue. In Table
\ref{catalog} we report new parameters for the first five objects from
this catalogue table with: DR1 positions; calculated proper motions
with errors; $G$ magnitudes and errors; number of observations in DR1;
\G~SourceID; $ \Delta G = G - G_{\rm est}$; number of DR1 entries
within 3\,arcsec and a flag that indicates if the calculated proper
motion was within 20\% of the published or estimated value. The
published catalogue also has other literature information such as 2MASS
and WISE magnitudes for each entry.

\section{Common proper motion LT dwarfs and DR1 stars}
\label{section:cpm}
\subsection{The Tycho-\G~astrometric subset}

The \G~DR1 included a subset of more than 2 million objects that incorporated
earlier positional information to find parallaxes and proper motions called
the Tycho-\G~astrometric solution
\citep[TGAS;][]{2015A&A...574A.115M,2016A&A...595A...2G}.  We selected CPM
pairs by cross-matching our catalogue of known LT dwarfs with the TGAS subset.
For the input LT dwarfs we used measured parallaxes from the literature,
complemented with spectrophotometric distances estimated using the adopted UCD
spectral types and near-infrared magnitudes.  To estimate their
spectrophotometric distance we used the polynomial relations presented in
\citet{2012ApJS..201...19D}, with the measured 2MASS $J$ magnitude, and, if
not available or the 2MASS value has a bad quality flag (Qflag = U), we use 
the MKO $J$ magnitude.

\subsection{Selection criteria}
\label{selection_criteria}

The starting CPM candidate list was generated from finding all TGAS stars within 2\,deg of
our input LT dwarfs and applying the following criteria:

\begin{itemize}
	\item $\mu$~$>$~100~mas~yr$^{-1}$
	\item $\Delta\mu_\alpha \cos{\delta} <\,$20~mas~yr$^{-1}$ and $\Delta\mu_\delta\,<\,$20~mas~yr$^{-1}$
\end{itemize}

\noindent where $\mu$ is the total proper motion,
$\Delta\mu_\alpha\cos{\delta}, \Delta\mu_\delta$ are the difference between
the proper motion components of the UCD and the TGAS star. All selected TGAS objects are close so we do not need
to invoke inference techniques to find distances \citep[e.g.][]{2015PASP..127..994B},
but use the simple inverse of the parallax as the estimated distance and as
its error a proportion equal to the relative error of the parallax. We then
calculated a chance alignment probability for each system following the method
described in Marocco et al. (2017). 

	
The selection criteria requires the objects to have relatively high proper
motions and the probability of having two objects with such high proper
motions in a limited area is already small. For each candidate pair we used
the sample of all TGAS field stars in a radius of 2\,deg from the UCD to
determine the distance and proper motion distribution of the field
population. The distance and proper motion distribution were treated as a
probability density function, which we reconstructed using a kernel density
estimation. We then draw 10,000 samples of stars from the reconstructed
probability density function, and determined how many ``mimics'' of our system
were generated. We considered as a mimic of our CPM system any star within
3$\sigma$ of the distance and proper motion of our selected primaries. The
chance alignment probability was assumed to be the number of mimics divided by
10,000. If this probability was below 6\,$\times$10$^{-5}$, equivalent to a
4$\sigma$ level, we consider the pair to be a ``robust'' common proper motion
system. Systems with larger chance alignment probability were ruled out.

\begin{figure*}
\centering
\includegraphics[width=0.45\textwidth]{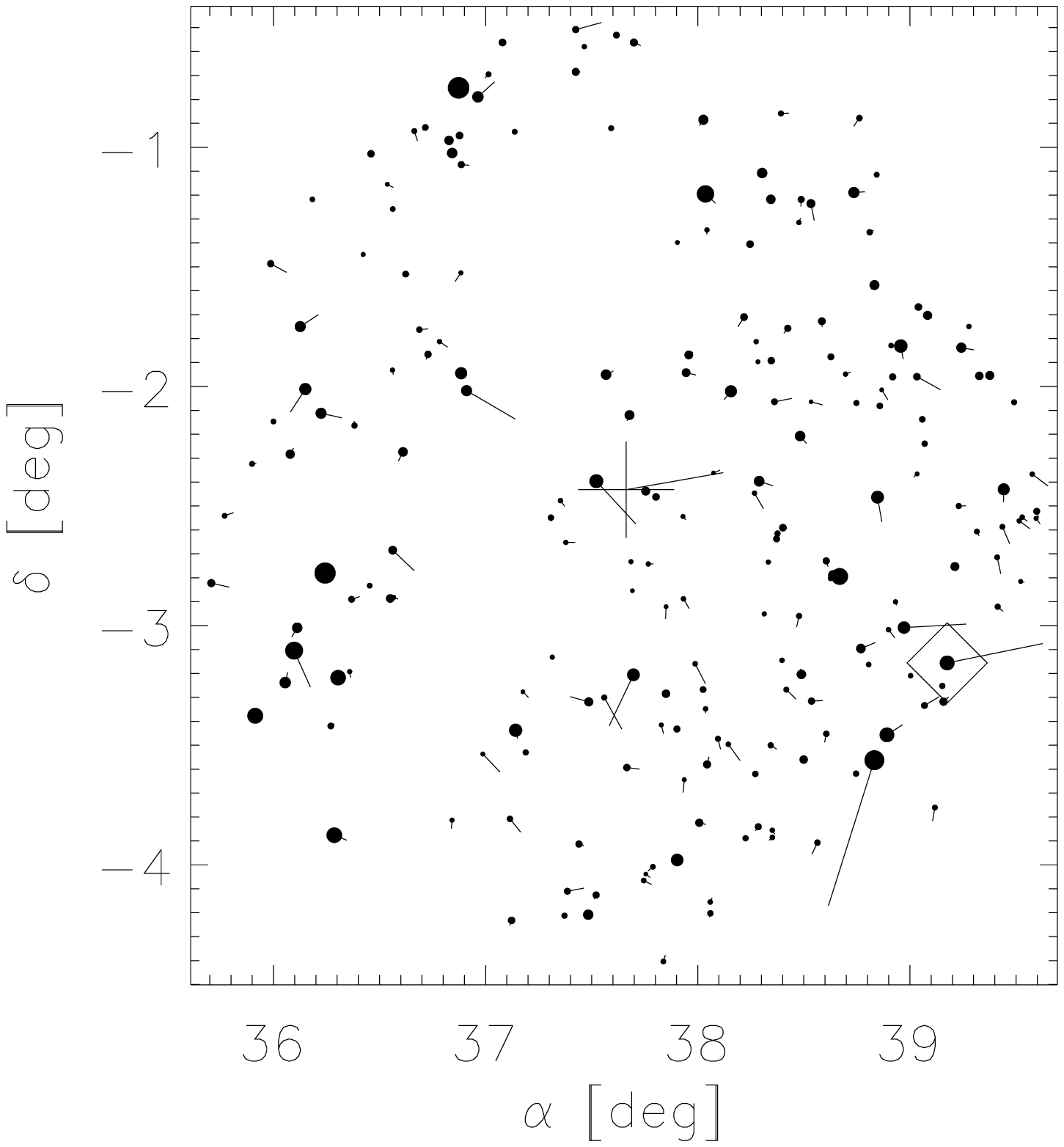}
\includegraphics[width=0.45\textwidth]{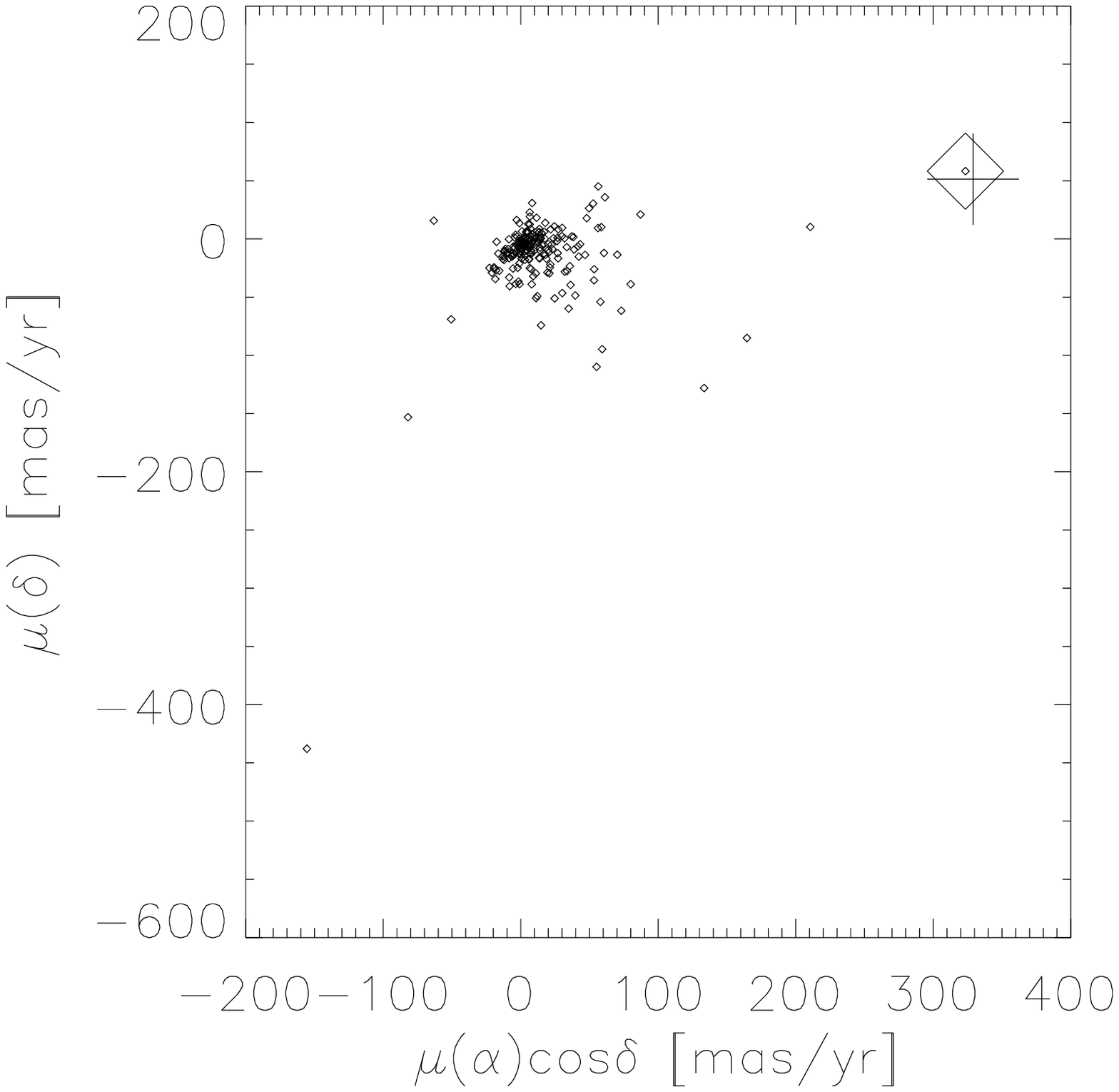}
\caption{A probable wide CPM: TGAS objects within 2~degrees of the L8
  J0230-0225.  In the left panel we show the TGAS on-sky positions with proper
  motions indicated by the vectors and the size of the symbols indicating the
  magnitude. The large diamond encloses the TGAS candidate CPM HIP 12158 and
  the plus sign indicates the position of J0230-0225.  In the right panel we
  show a vector point diagram for all objects again with the cross indicating
  J0230-0225, the small symbols are the proper motions of the TGAS objects and
  the large diamond encloses HIP 12158.\label{cpmplots1} }
\includegraphics[width=0.45\textwidth]{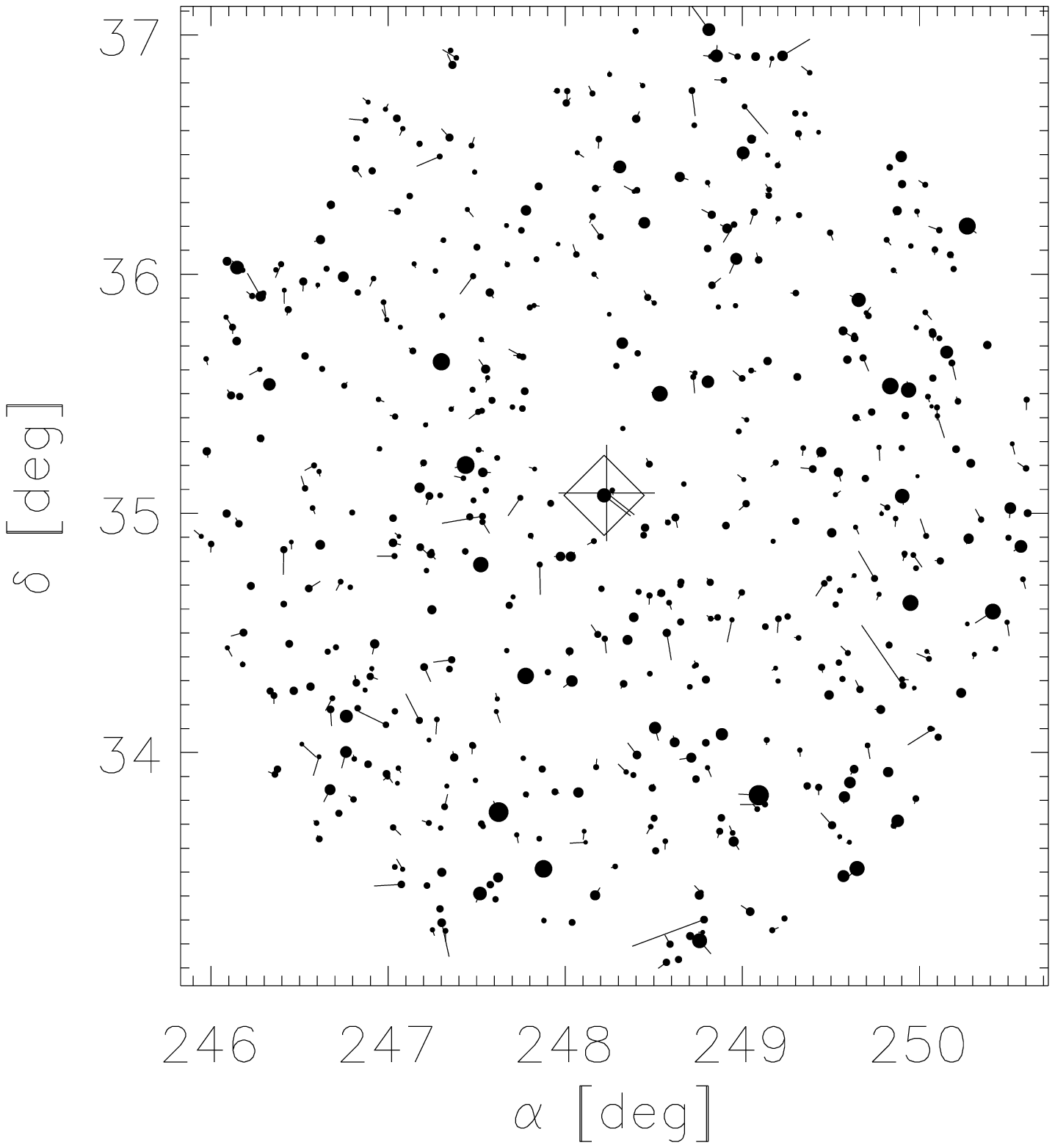}
\includegraphics[width=0.45\textwidth]{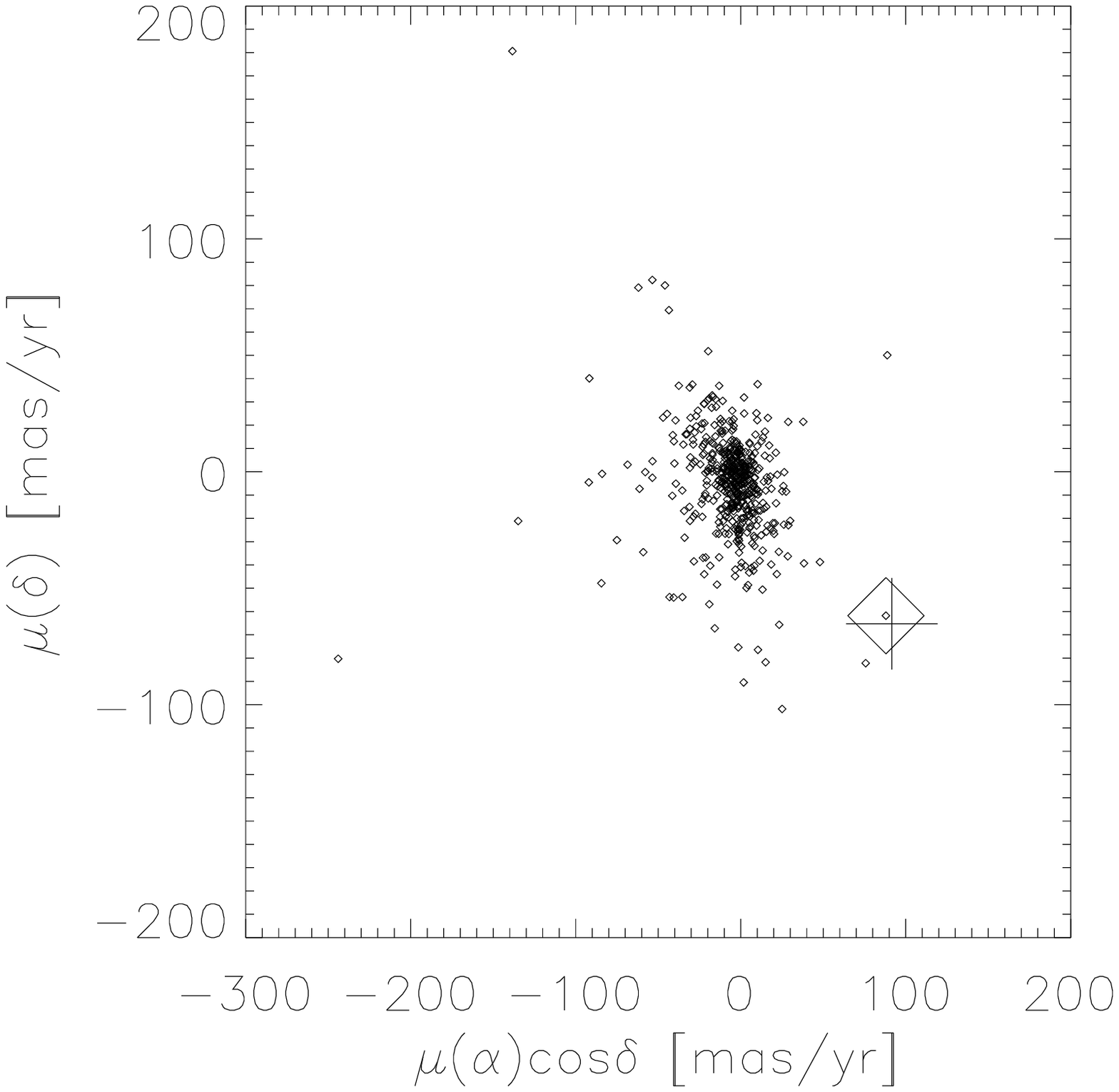}
\caption{A probable binary: TGAS objects within 2~degrees of the L0.5
  J1632+3505.  The symbol meanings are as figure 8 however, the TGAS object,
  TYC 2587-1547-1, overlaps the L dwarf position in this
  example.\label{cpmplots2} }
\end{figure*}
%
%

\begin{figure}
\includegraphics[width=0.49\textwidth]{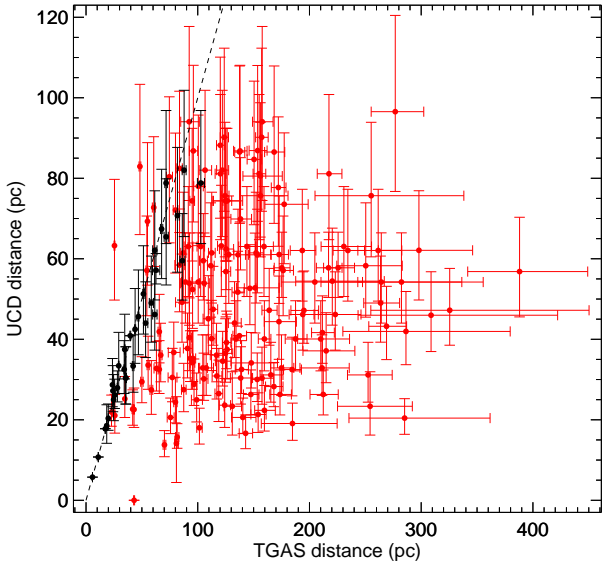}
\caption{A comparison between the measured TGAS distance of the primaries vs
  distances to their candidate UCD companions for the full sample
  passing our angular separation and proper motion selection. Systems selected
  as common-distance are plotted in black, while those rejected are plotted in
  red. \label{dist_comparison}} 
\end{figure}

\begin{figure}
\includegraphics[width=0.49\textwidth]{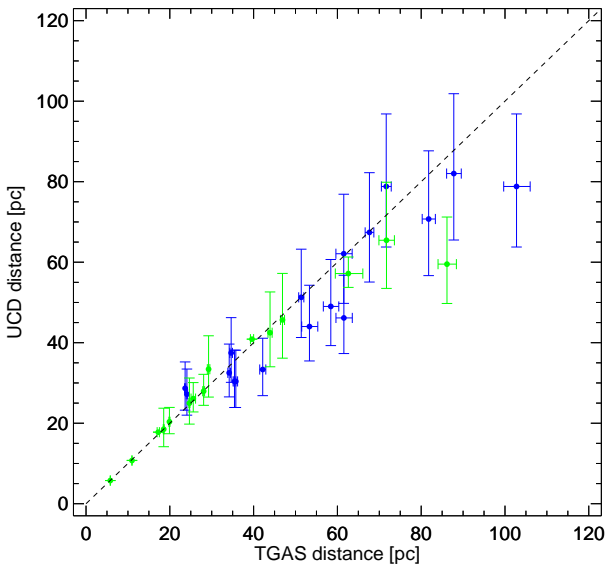}
\caption{A comparison of TGAS and UCD distances for common-distance selected
  systems. Systems where the UCD has a measured parallax are plotted in green,
  while those with only spectrophotometric distance estimates are plotted in
  blue. \label{dist_comparison2}}
\end{figure}

\subsection{LT dwarfs and \G~CPM system catalogue}
This selection yielded a sample of \TOTCPMS~CPM pair candidates. We
compiled a list of known binary and CPM systems by combining the objects and
list from the following publications: \cite{2001AJ....122.3466M},
\cite{2014ApJ...792..119D}, \cite{2014MNRAS.445.3694D},
\cite{2015AJ....150...57D}, \cite{2015ApJ...804...96G},
\cite{2015MNRAS.454.4476S}, \cite{2016A&A...587A..51S},
\cite{2016ApJS..224...36K}, \cite{2016arXiv161208873G}, and
\cite{2017arXiv170103104D}.  
Of the \TOTCPMS~CPM pair candidates \OLDCPMS~were previously known and the
remaining \NEWCPMS, 
listed in Table~\ref{summary_table}, are presented here for the
first time.  The majority of new wide systems presented here are not
physically bound pairs, but the low chance alignment
probabilities we interpret as an indication of common origin. Intrinsically
wide binaries and multiple systems can in fact become unbound due to Galactic
tides and close encounters
\citep[e.g.][]{2016MNRAS.463.2958V,2016MNRAS.459.4499E}, and their ejecta
would represent a new, as yet unexplored pool of benchmark systems
\citep{2006MNRAS.368.1281P,2016csss.confE.137Y}. In Figs. \ref{cpmplots1}
and \ref{cpmplots2} we plot celestial and proper motion distributions of
example unbound (J0230-0225) and bound (J1632+3505) CPM pairs discussed later.

%
%
One key element in our selection process is the requirement of common distance
between the main sequence TGAS star and its potential companion. We show in
Figure~\ref{dist_comparison} a comparison between the measured astrometric
distance to the primaries in our CPM pairs, against the distance (astrometric
or spectrophotometric) to their potential companions. Common distance systems
are highlighted in black. Uncertainties on the spectrophotometric distance
dominate, and at larger distance this results in a much larger scatter
around the one-to-one correspondence line.
Pairs that passed our angular separation constraint, but were rejected
by the common-distance cut, consist of a foreground UCD matched to a 
background star. In Figure~\ref{dist_comparison2} we
plot only those systems that we select as having common distance, with those
UCDs with measured parallaxes highlighted in green. As expected,
systems with astrometric measurements are much closer to the one-to-one
correspondence line than those with spectrophotometric distance estimates
only. The UCD spectroscopic distances tend to be underestimated compared to the
TGAS parallactic distances, e.g. the UCD is brighter than the spectral type
indicates. This is as expected from unresolved binarity or a Malmquist 
bias like effect as our input sample is probably biased to the brighter 
examples of a given spectral class bin. 

%
%

\begin{table*}
\caption{The common-distance, common-proper-motion pair candidates identified
  here. A machine readable version of this table is available online. \label{summary_table}}
\label{tab:landscape}
\begin{tabular}{l c c c c c c c c}
\hline
Name  & $\alpha$ &$ \delta$      & $\mu_\alpha\cos{\delta}$ &  $\mu_\delta$
& $d$    & PDF  & Proj. Sep. & SpT \\
      & [deg] & [deg]     & [mas yr$^{-1}$]    & [mas yr$^{-1}$]        &
[pc]& & [10$^3$\,au] & \\
\hline
   HIP 10346    &  33.31768733 & -59.56700153 &    126.5$\pm$ 0.1 &     -8.3$\pm$ 0.1 &   58.4$\pm$ 1.8 & 2.2e-05 & 400\\
     J0223-5815 &  35.97766900 & -58.25187300 &    134.0$\pm$10.0 &      5.0$\pm$19.0 &   49.0$\pm$10.0 &  & \\
\hline
   HIP 12158    &  39.17535990 &  -3.15589420 &    323.4$\pm$ 0.1 &     58.2$\pm$ 0.1 &   24.1$\pm$ 0.2 & 3.6e-08 & 145\\
     J0230-0225 &  37.66214300 &  -2.43167270 &    329.0$\pm$16.8 &     51.3$\pm$14.9 &   27.0$\pm$ 6.0 &  & \\
\hline
TYC 146-1101-1  &  97.54755003 &   0.87648584 &     68.4$\pm$ 2.9 &   -105.7$\pm$ 2.4 &   67.6$\pm$ 1.0 & 3.3e-06 & 252\\
     J0626+0029 &  96.58839400 &   0.49281806 &     84.0$\pm$15.0 &    -92.0$\pm$15.0 &   67.0$\pm$14.0 &  & \\
\hline
   HIP 38492    & 118.24796721 &  22.55612499 &    -85.9$\pm$ 0.2 &    -61.4$\pm$ 0.1 &   34.2$\pm$ 0.3 & 3.0e-07 & 157\\
     J0758+2225 & 119.62429000 &  22.42408300 &   -105.0$\pm$ 8.0 &    -62.8$\pm$ 8.2 &   33.0$\pm$ 8.0 &  & \\
\hline
TYC 230-109-1   & 139.99273886 &   4.99944728 &    -80.0$\pm$ 1.2 &    -40.2$\pm$ 1.1 &   42.2$\pm$ 0.7 & 1.5e-07 & 178\\
     J0915+0531 & 138.93388000 &   5.51780560 &    -95.0$\pm$ 5.5 &    -57.7$\pm$ 4.4 &   33.0$\pm$ 6.0 &  & \\
\hline
TYC 2504-466-1  & 145.55216670 &  33.92992123 &   -103.7$\pm$ 3.8 &    -67.3$\pm$ 1.6 &   61.5$\pm$ 1.9 & 1.1e-06 & 156\\
     J0939+3412 & 144.77713000 &  34.21596100 &   -107.1$\pm$10.4 &    -64.3$\pm$12.6 &   62.0$\pm$12.0 &  & \\
\hline
   HIP 47704    & 145.89719170 &  10.51834984 &     37.5$\pm$ 0.1 &   -124.9$\pm$ 0.1 &   71.7$\pm$ 1.2 & 2.5e-06 & 211\\
     J0943+0942 & 145.95667000 &   9.70094440 &     45.4$\pm$10.9 &   -119.9$\pm$ 8.8 &   79.0$\pm$15.0 &  & \\
\hline
TYC 824-423-1   & 145.44963031 &  11.17517624 &     64.6$\pm$ 0.9 &   -111.1$\pm$ 0.7 &  102.8$\pm$ 3.2 & 8.2e-06 & 576\\
     J0943+0942 & 145.95667000 &   9.70094440 &     45.4$\pm$10.9 &   -119.9$\pm$ 8.8 &   79.0$\pm$15.0 &  & \\
\hline
   HIP 57734    & 177.58699124 &  10.06723706 &    -89.9$\pm$ 0.1 &    -16.5$\pm$ 0.0 &   86.2$\pm$ 2.2 & 1.2e-06 &  77\\
     J1150+0949 & 177.66163000 &   9.82858330 &   -107.6$\pm$17.1 &    -31.9$\pm$ 4.5 &   60.0$\pm$27.0 &  & \\
\hline
   HIP 58241    & 179.18153043 & -32.26744260 &   -178.8$\pm$ 0.6 &     -7.1$\pm$ 0.3 &   35.4$\pm$ 0.3 & 5.7e-06 & 229\\
     J1154-3400 & 178.67596000 & -34.01084900 &   -161.0$\pm$13.0 &      4.0$\pm$15.0 &   30.0$\pm$ 6.0 &  & \\
\hline
   HIP 58240    & 179.17544959 & -32.26819149 &   -172.0$\pm$ 0.4 &     -8.3$\pm$ 0.3 &   35.8$\pm$ 0.4 & 4.0e-06 & 231\\
     J1154-3400 & 178.67596000 & -34.01084900 &   -161.0$\pm$13.0 &      4.0$\pm$15.0 &   30.0$\pm$ 6.0 &  & \\
\hline
   HIP 59887    & 184.22818632 &  37.48407627 &   -107.9$\pm$ 0.1 &     -2.0$\pm$ 0.1 &   87.8$\pm$ 1.7 & 9.0e-06 & 153\\
     J1214+3721 & 183.64024000 &  37.35327100 &   -122.6$\pm$10.6 &     15.7$\pm$13.4 &   82.0$\pm$17.0 &  & \\
\hline
   HIP 62350    & 191.64195696 &  11.37849569 &   -112.4$\pm$ 0.1 &     -0.5$\pm$ 0.0 &   61.5$\pm$ 1.9 & 1.1e-05 & 286\\
     J1244+1232 & 191.05429000 &  12.53363900 &   -104.8$\pm$ 8.6 &      4.5$\pm$ 7.3 &   46.0$\pm$ 8.0 &  & \\
\hline
TYC 2587-1547-1 & 248.21928595 &  35.07510874 &     88.0$\pm$ 0.5 &    -61.8$\pm$ 0.5 &   34.6$\pm$ 0.3 & 1.4e-10 &   2\\
     J1632+3505 & 248.23375000 &  35.08545700 &     91.6$\pm$ 9.7 &    -65.3$\pm$11.9 &   37.0$\pm$ 8.0 &  & \\
\hline
  HIP 101880    & 309.67956167 & -43.73289384 &    235.1$\pm$ 0.2 &   -371.5$\pm$ 0.1 &   51.4$\pm$ 0.6 & 2.1e-10 & 270\\
     J2037-4216 & 309.46379000 & -42.27922200 &    229.0$\pm$10.0 &   -391.0$\pm$10.0 &   51.0$\pm$10.0 &  & \\
\hline

\end{tabular}
\end{table*}

\subsection{Selected CPM systems}

Extrapolating the simulations in Marocco et al. (2017) we predicted that the
number of confirmable LT binary systems for the TGAS subset of \G~DR1 is more
than 100, while the number of unbound, but still common proper motion systems,
is significantly higher. The procedures modelled in Marocco et al. (2017) did
not include moving groups and disintegrating clusters as our knowledge of
these systems is still in its infancy, hence the total number of CPM systems
between LT dwarfs and the DR1 TGAS subset is probably many hundreds. The
ongoing large scale infra-red surveys will provide a complete list of nearby
LTs and at that point a comparison to \G~results will also allow us to
constrain many of the uncertain factors used in the Marocco et al. (2017)
work.

For the illustration of the diverse characteristics and possible uses of the
CPM systems presented here it is useful to consider a few of the systems
individually:
\begin{itemize}
\item J0230--0225 is an L8 with a peculiar spectrum \citep{Thompson2013}
  that we have associated to HIP~12158 ({FT~Cet}) a K1V star that has
  been indicated as a member of the Hyades Moving Group         
  \citep{2012A&A...547A..13T}. If coeval with the Hyades it will have an age
  of between 0.4-1.0\,Gyr and any interpretation of the spectra peculiarities
  will have to take that into consideration.
\item J0915+0531 is a T7 associated to TYC~230--109--1 ({HD~80462}),
  the primary of a visual binary system of two mid-G-type stars separated by
  10\arcsec and discovered by F.\,G.\,W. Struve in the early 19th century
  \citep{2001AJ....122.3466M}. For such a binary system \G~will produce
  precise astrometry and high resolution RVS spectra that will provide
  a significant improvement on their astrophysical parameters which in turn can
  be used to constrain the UCD if the companionship is confirmed.

\item J1154-3400 \citep{2008AJ....135..785W} is an L0
  was proposed as a candidate member of of the Argus
  Association \citep{2015ApJS..219...33G}, which is a
  young 30–50 Myr system, so if this UCD is a member it
  is in an age regime where the radius is rapidly
  changing \citep{2002A&A...382..563B}. Further work in
  \citet{2016ApJS..225...10F} evidence that it as a
  very diffcult case with moderate to high
  probabilities of being in various kinematic
  groups. We associate this dwarf to the primary HIP
  58240 ({HD~103742}), in a G4V+G3V binary system with
  HIP 58241 ({HD~103743}). Adopting the precise TGAS
  astrometry of the binary system it did not register
  as a candidate member of any moving group in either
  the BANYAN II \citep{2014ApJ...783..121G} or LACEwING
  \cite{2017AJ....153...95R} tools for estimate
  probabilities of candidate objects to nearby
  kinematic groups. BANYAN II did indicate the system
  maybe young (probability of being a young field
  object 47\% against old field of 53\%). 

  Adopting our proper motion of J1154-3400 and it's
  parallax from \citet{2009AJ....137....1F} in BANYAN
  II we find there is an indication it maybe a member
  of Argus (probability 53\%) or the TW Hydrae
  Association (probability 10\%) while LACEwING gives
  zero probabilities for all moving groups. If the
  connection to the HIP 58240/58241 system is confirmed
  it would be hard for us to also conclude it is part
  of the nearby moving groups. We await future releases
  of the \G~results to resolve these conflicting
  indications.

\item J1632+3505 (L0.5) and TYC~2587--1547--1 (HD~149361, K0\,V) are separated
  by $\rho$ = 57\,arcsec at position angle $\theta$ = 49\,deg.  At a
  heliocentric distance of 34.6$\pm$0.3\,pc this translates into a projected
  physical separation of only 1960\,au and a gravitational potential
  energy of the order of $-10^{-35}$\,J, between 40 and 300 times larger in
  absolute value than the most fragile bound systems
  known \citep{2009A&A...507..251C}.  The relatively short projected physical
  separation, large absolute potential energy, and similarity of recalculated
  proper motions of both primary and secondary with \G~led us to classify the
  pair as the only bound system in our sample.The primary star is at the
  brighter end of the \G~magnitude range ($G$=7.97\,mag) while J1632+3505 is
  at the faint end ($G$=19.18\,mag), so the consistency of the two
  \G~distances will be testing both noise and photon limited astrometric
  results. The \G~spectroscopic observations of the primary will lead to
  astrophysical parameters that can be used to constrain those of the
  secondary.
\item J0943+0942 is a T4.5 that is found to be a CPM companion candidate of two TGAS
  stars ( HIP 47704 and TYC 824-423-1 ). However, from the more precise proper
  motions of the two stars they would
  not be considered CPM companions. This highlights a weakness of our
  procedure, we calculate a probability but there will be false positives.
  This is one of the faintest objects that we found CPM pairs for and like most
  UCDs in these systems \G~will not detect them, however, we expect future
  infrared and deep optical surveys to allow us to improve the proper motion
  of all UCDs.
\end{itemize}


\section{An examination of GUCDS magnitude, colour and proper motion relations}

The \G~$G$ is a new passband from 330\,nm to 1050\,nm with transmission
peaking around 600\,nm and dropping to 10\% at
970\,nm\footnote{\url{http://www.cosmos.esa.int/web/gaia/science-performance}}.
The $G$ magnitude represents a new resource both in terms of homogeneity and
wavelength coverage, albeit with possibly limited diagnostic ability for short
baseline colours due to its very wide spectral passband. It will help to
constrain the spectral energy distribution of LT dwarfs across the whole of
the sky in the optical. As well as this $G$ magnitude, the second \G~data
release will provide both a blue and a red magnitude ($G_{BP}$ and $G_{RP}$,
see Fig. \ref{passbands}), the application of which to UCDs has been discussed
at length in Sarro et al. (2012). In this section we examine the locus of UCD
objects in relations between magnitude, colour and proper motion with a focus
on those related to the \G~$G$ band.

\label{section:results}
\begin{figure}
\centering
\includegraphics[width=0.49\textwidth]{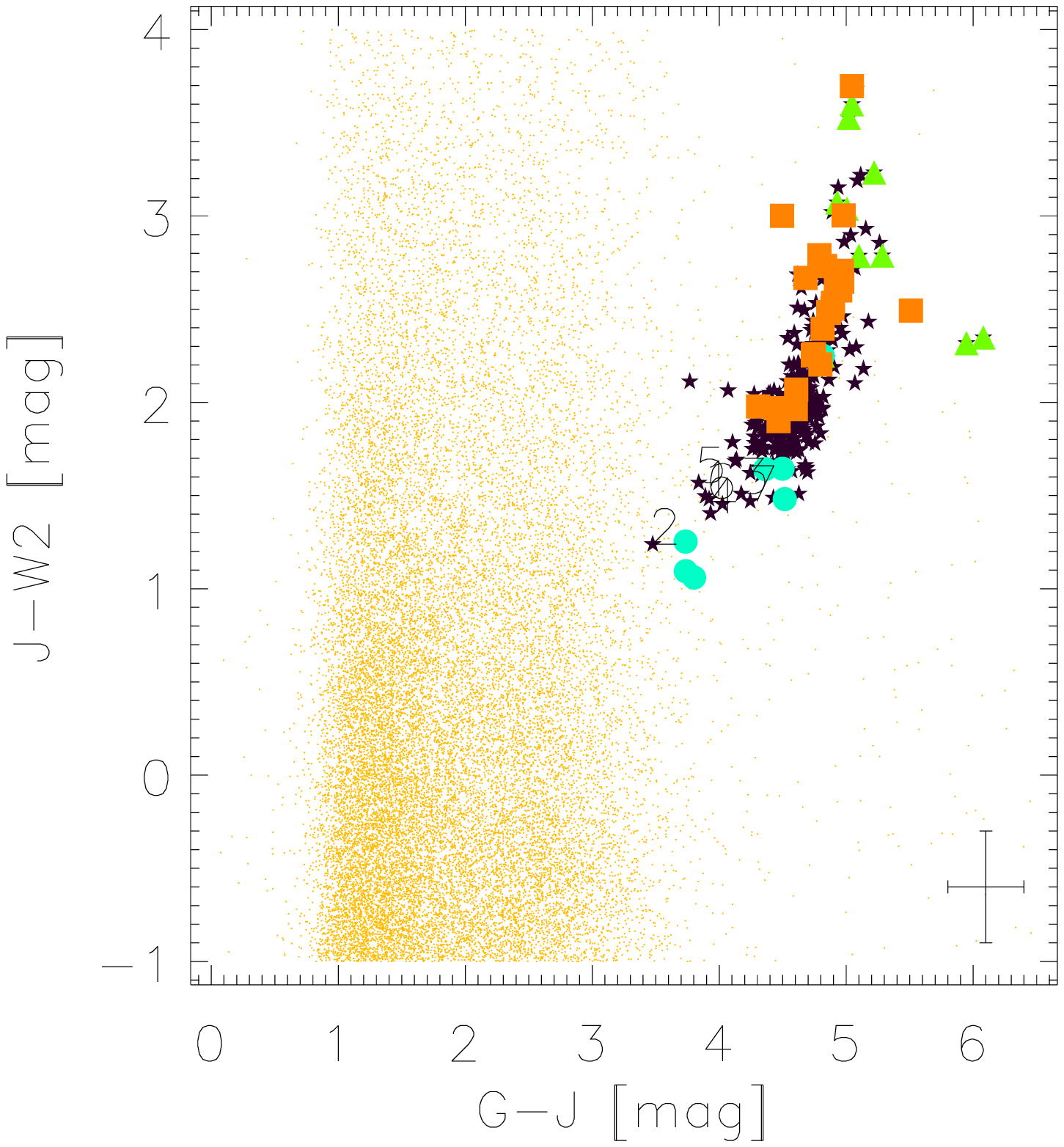}
\includegraphics[width=0.49\textwidth]{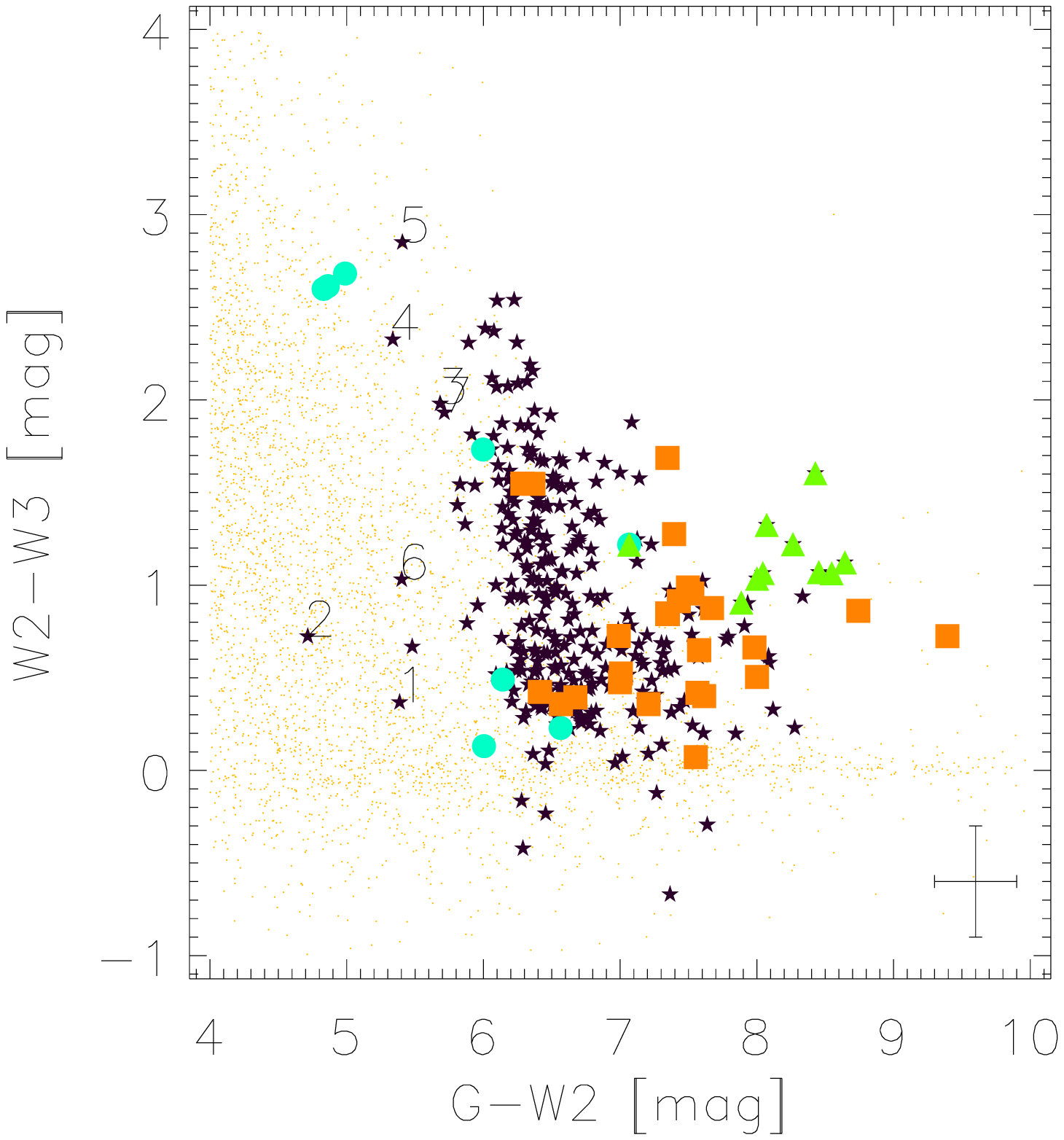}
\caption{Colour-colour diagrams for $J-W2$ vs. $G-J$ and a zoom on the
  LT region of the $W2-W3$ vs. $G-W2$. Filled circles are subdwarfs, squares are young dwarfs, triangles are
objects later than L7, stars are other L0--6 dwarfs, and yellow points are anonymous sources within 2\,arcmin of
each LT dwarf. Typical error bars are shown. The seven outliers J0109-4954, J0133-6314,
  J1116+6037, J1245+4902, J1250+4418, J1251+6243 and J1333+1509 are labelled 1
  through 7 respectively. Additional colour-colour
diagrams are found in Appendix~A.\label{COL1}}
\label{COL1}
\end{figure}

\subsection{Colour - colour relations}

In the optical wavebands, LT dwarfs only have homogeneous magnitude estimates
for parts of the sky \citep[e.g. see][]{1999AA...349..236E,
  2014ApJS..211...17A}, while in the infrared most have homogeneous magnitudes
from the 2MASS and AllWISE surveys. This is also where these objects are the
brightest, and combined with $G$ produce a very long colour baseline, so here
we examine the \G~and 2MASS/WISE infrared relations. The 2MASS is 99\,\%
complete to
$J=15.8$\,mag\footnote{\url{http://www.ipac.caltech.edu/2mass/overview/about2mass.html}}
and, from Figure \ref{sptvsGmJ1}, the nominal $G-J$ offset of an L0 is
4.5\,mag; we therefore expect 2MASS to be complete for all \G~objects with
spectral types L0 and later (as they are redder) to $G=20.3$\,mag. The
magnitude limit of AllWISE is $W2=15.7$\,mag and the colour offset is greater,
so all \G~LT dwarfs are expected to have an ALLWISE $W2$ detection.

In Fig. \ref{COL1} we plot $J-W2$ as a function of $G-J$ and zoomed in
on the LT portion of the $W2-W3$ vs. $G-W2$ graph. In Appendix~A
we plot the LT region in a series of combinations of $G$
with 2MASS and AllWISE magnitudes. The $G$ related blue colour is always
on the X axis and the redder magnitude combination on the Y axis. We did
not include $W4$ as it is generally not sensitive enough to detect these
objects.  To highlight particular populations, we have plotted different
symbols for the young or subdwarfs from the comments of Table 1, and the
\NGTL~objects with spectral types later than L7.

\begin{table*}
\begin{center}
  \caption[]{Probable late M dwarfs identified in the
colour-colour and colour-spectral type diagrams. }
  \label{outliers}
\begin{tabularx}{\textwidth}{llX}
    \hline
    \hline
    \noalign{\smallskip}
Long Name   & Opt SpT    &  Remarks    \\
Short Name  & NIR SpT    &            \\ 
Plot label  & $\Delta G$ &        \\ 
    \noalign{\smallskip}
    \hline
    \noalign{\smallskip}
\vtop{\hbox{\strut SSSPM~J0109--4955     }\hbox{\strut J0109-4954}\hbox{\strut
    1}} & \vtop{\hbox{\strut M8$^1$ }\hbox{\strut L1$^2$}\hbox{\strut -0.45}}
& This object is bluer in all colours than we would expect for an M8 but
further spectral observations are needed to clarify its spectral type.  \\
\hline
\vtop{\hbox{\strut SSSPM~J0134--6315     }\hbox{\strut J0133-6314}\hbox{\strut
    2}} & \vtop{\hbox{\strut ...    }\hbox{\strut L0$^2$}\hbox{\strut -1.02}}
& This is the bluest LT dwarf  in the majority of colour-colour plots. It has
been classified as early as an M5 \citep{2005AA...436..853L} and a X-Shooter
spectra  
is a best fit with an M6 template. The $W3$ magnitude is an upper limit which explain its outlier position in the $W3$ colour combinations.  \\ 
\hline 
\vtop{\hbox{\strut 2MASS~11164800+6037309}\hbox{\strut J1116+6037}\hbox{\strut 3}} & \vtop{\hbox{\strut L0$^3$ }\hbox{\strut ...   }\hbox{\strut -0.33}} & This is at the border of the majority of L0 colour loci.  Using the best fitting template procedure from \cite{2015MNRAS.449.3651M} on its published SDSS spectra we find it is a late M dwarf. \\ 
\hline 
\vtop{\hbox{\strut 2MASS~12455566+4902105}\hbox{\strut J1245+4902}\hbox{\strut
    4}\hbox{\strut 1}} & \vtop{\hbox{\strut L1$^3$ }\hbox{\strut
    M8$^4$}\hbox{\strut -0.65}} & This object has been classified as M8$^4$ in
the infrared and both L1$^3$ and M9 \citep{2011AJ....141...97W} in  the
optical -- both from SDSS spectra. It is a borderline M/L object.  \\
\hline 
\vtop{\hbox{\strut 2MASS~12504567+4418551}\hbox{\strut J1250+4418}\hbox{\strut
    5}} & \vtop{\hbox{\strut L0$^5$ }\hbox{\strut ...   }\hbox{\strut -0.66}}
& Very blue in many colour-colour plots and comparisons of the SDSS spectra
using the procedure from \cite{2015MNRAS.449.3651M}, the same spectra as used
by West et al. (2008) to find L0, we find the object to be a late M dwarf. The $W3$ magnitude is an upper limit and its extreme position in the W3 colour combinations indicates that it is significantly fainter that the published value.\\ 
\hline 
\vtop{\hbox{\strut 2MASS~12512841+6243108}\hbox{\strut J1251+6243}\hbox{\strut
    6}} & \vtop{\hbox{\strut M8V$^5$}\hbox{\strut L4$^4$}\hbox{\strut -0.90}}
& This object was listed by as an L4$^4$ but erroneously cited as coming from
\cite{2009AA...497..619Z}; \cite{2008AJ....135..785W} classified it as an
M8V. Our $G$ magnitude would be more consistent with the earlier type and we
think this is a case of object mis-identification and the actual object is an
M8V.\\ 
\hline 
\vtop{\hbox{\strut 2MASS~13331284+1509569}\hbox{\strut J1333+1509}\hbox{\strut
    7}} & \vtop{\hbox{\strut L0$^3$ }\hbox{\strut M8$^4$}\hbox{\strut -0.26}}
& This object has been classified as M8$^4$ in the infrared and both L0$^3$
and M9 \citep{2011AJ....141...97W} from the same SDSS spectra in the
optical. It is a borderline M/L object.  The $W3$ magnitude is an upper limit
and it is an outlier in the ALLWISE $W3$ colour combinations so it is probably
fainter.\\ 
\noalign{\smallskip} 
\hline
  \end{tabularx}
\end{center}
%
Spectral type references: 
$^1$\cite{2008AJ....136.1290R}, 
$^2$\cite{2005A&A...436..853L}, 
$^3$\cite{2010AJ....139.1808S}, 
$^4$\cite{2014APJ...794..143B}, 
$^5$\cite{2008AJ....135..785W}
\end{table*}

We find that the seven objects listed in Table \ref{outliers}, that are
not indicated as young or subdwarf objects, are outliers in most
colour-colour plots, and we have labelled them as 1 through 7
respectively on the figures. If these seven objects are classed as M/L
boundary dwarfs they would no longer be considered outliers as their
position is with the majority of M dwarfs, that dominate the background
objects.


From Fig. \ref{COL1} and the colour-colour plots in Appendix~A
we note the \G--2MASS colours differentiate better the full sample from the
background sources, while the subdwarf, young objects and late LT objects are
differentiated better when the AllWISE bands are used.  This variation in
properties  make it possible to photometrically isolate both the general
LT dwarfs and also the differing sub-populations.


\begin{figure}
\centering
\includegraphics[width=0.49\textwidth]{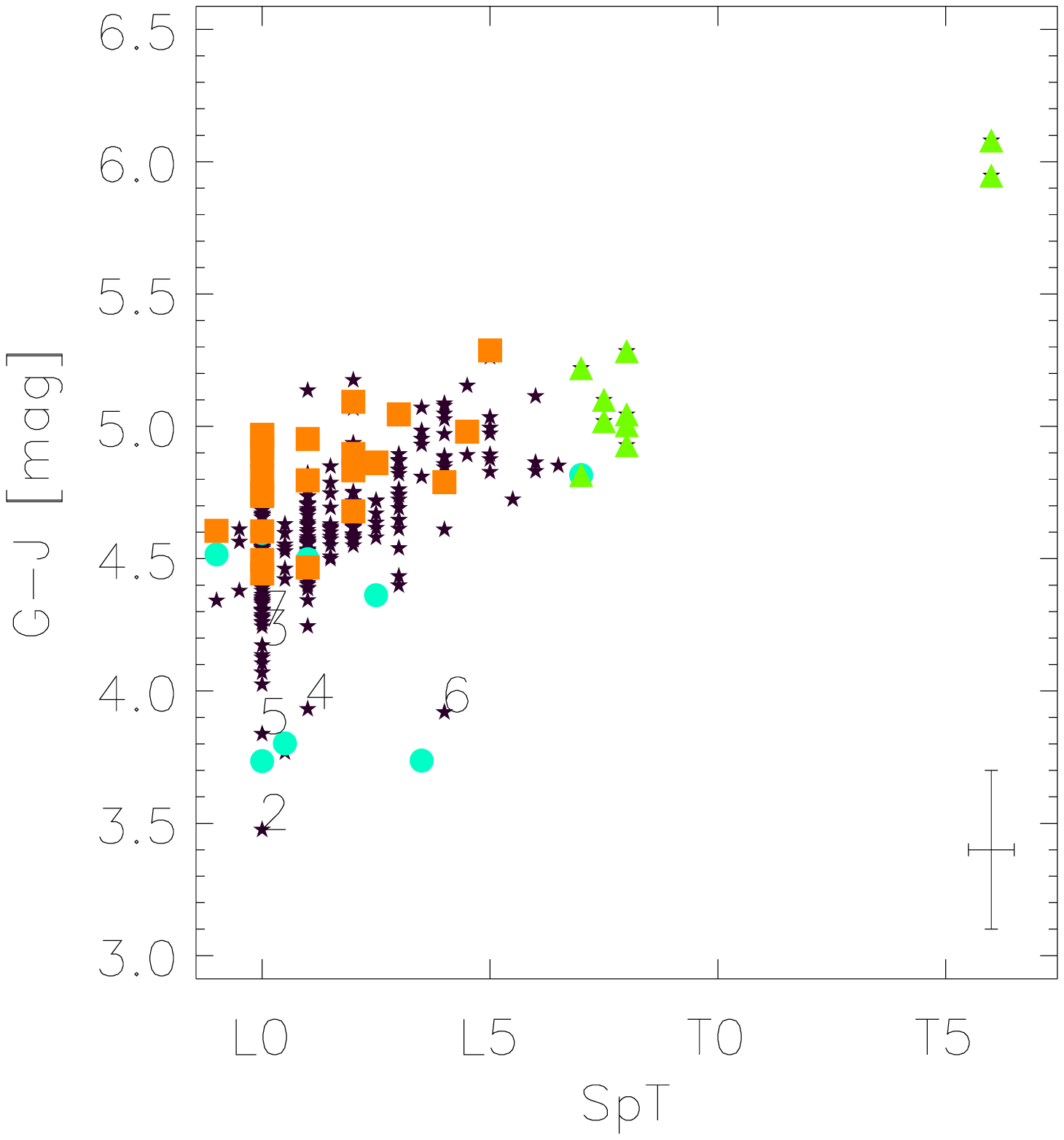}
\includegraphics[width=0.49\textwidth]{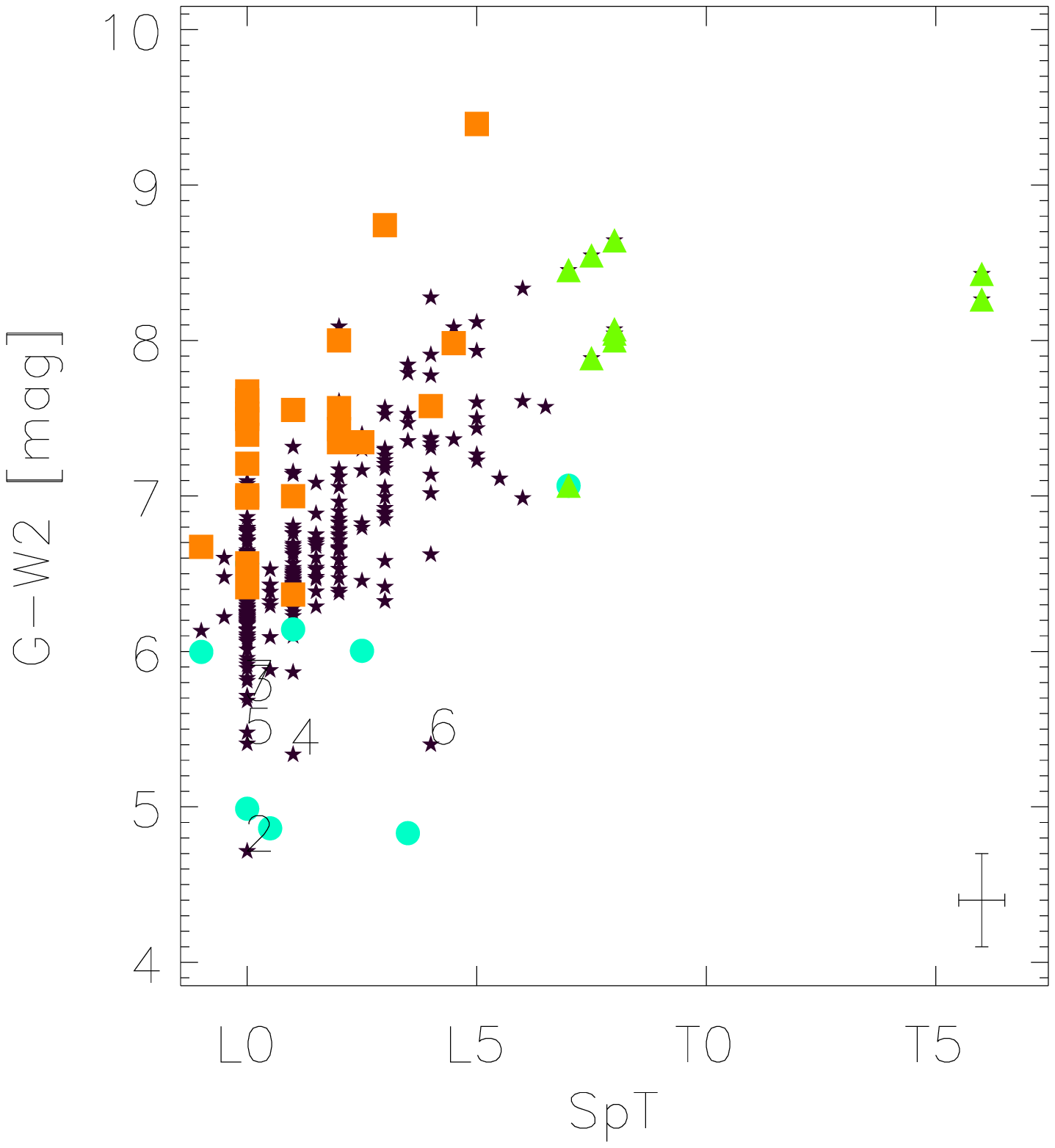}
\caption{Colour-spectral type diagrams for $G-J$ and $G-W2$. Symbols
  have the same meaning as Fig.  \ref{COL1}. The four outliers
  J0133-6314, J1245+4902, J1250+4418, J1251+6243 are labelled 1 -- 4
  respectively.  Additional colour-spectral type diagrams are found in
  Appendix~A.  }
\label{SPT1}
\end{figure}

\subsection{Spectral type and colour relations}

\begin{table}
\begin{center}
  \caption[]{Mean colours and standard deviation for all spectral
    type bins from L0 to L5 where there were at least 3 entries.}
  \label{mediancolours}
  \begin{tabular}{l cc}
    \hline
    \hline
    \noalign{\smallskip}
    Optical & $< G -J >$  & N     \\
     SpT    & [mag] & \\ 
    \noalign{\smallskip}
    \hline
    \noalign{\smallskip}
 L0   &  4.52 $\pm$ 0.19 &  126\\ 
 L0.5 &  4.53 $\pm$ 0.07 &    7\\ 
 L1   &  4.58 $\pm$ 0.13 &   55\\ 
 L1.5 &  4.63 $\pm$ 0.10 &   14\\ 
 L2   &  4.70 $\pm$ 0.15 &   29\\ 
 L2.5 &  4.65 $\pm$ 0.14 &    8\\ 
 L3   &  4.72 $\pm$ 0.17 &   17\\ 
 L3.5 &  4.95 $\pm$ 0.09 &    5\\ 
 L4   &  4.92 $\pm$ 0.14 &   11\\ 
 L4.5 &  5.01 $\pm$ 0.13 &    3\\ 
 L5   &  5.02 $\pm$ 0.17 &    8\\ 
    \noalign{\smallskip}
    \hline
  \end{tabular}
\end{center}
\end{table}

Using the  \NCAT~candidates with DR1 photometry, we can recalibrate
the $G - J$ as a function of SpT relation, which was used earlier to
provide an estimated $G$ value from spectral type and $J$
magnitude. In Table \ref{mediancolours} we report the mean colours,
standard deviation and number of entries for all spectral bins with
more than three objects.

In Fig. \ref{SPT1} we plot the $G-J$ and $G-W2$ with spectral type
graphs, and in Appendix~A 
we plot all of the $G$ based
colours with spectral types. In addition to estimating magnitudes, these
relations are useful for the identification of outlier objects, as they
are reasonably monotonic for the early L dwarfs. The onset of $Ks$ band
suppression due to the atomic and molecular absorption, methane in
particular \citep{1998ApJ...502..932O, 2000ApJ...541L..75N,
  2005ApJ...623.1115C}, and collision induced absorption
\citep{2001RvMP...73..719B, 2005ARAA..43..195K}, can be seen by the
position of the two T6 objects compared to the main bulk of the targets
in the $G-J$ and $G-K$ graphs. Young objects tend also to be redder in
the AllWISE colours, perhaps due to circum(sub)stellar discs or enhanced
dust formation due to low gravity \citep[e.g.][]{2010ApJ...715.1408Z}.


\begin{figure}
\centering
\includegraphics[width=0.52\textwidth]{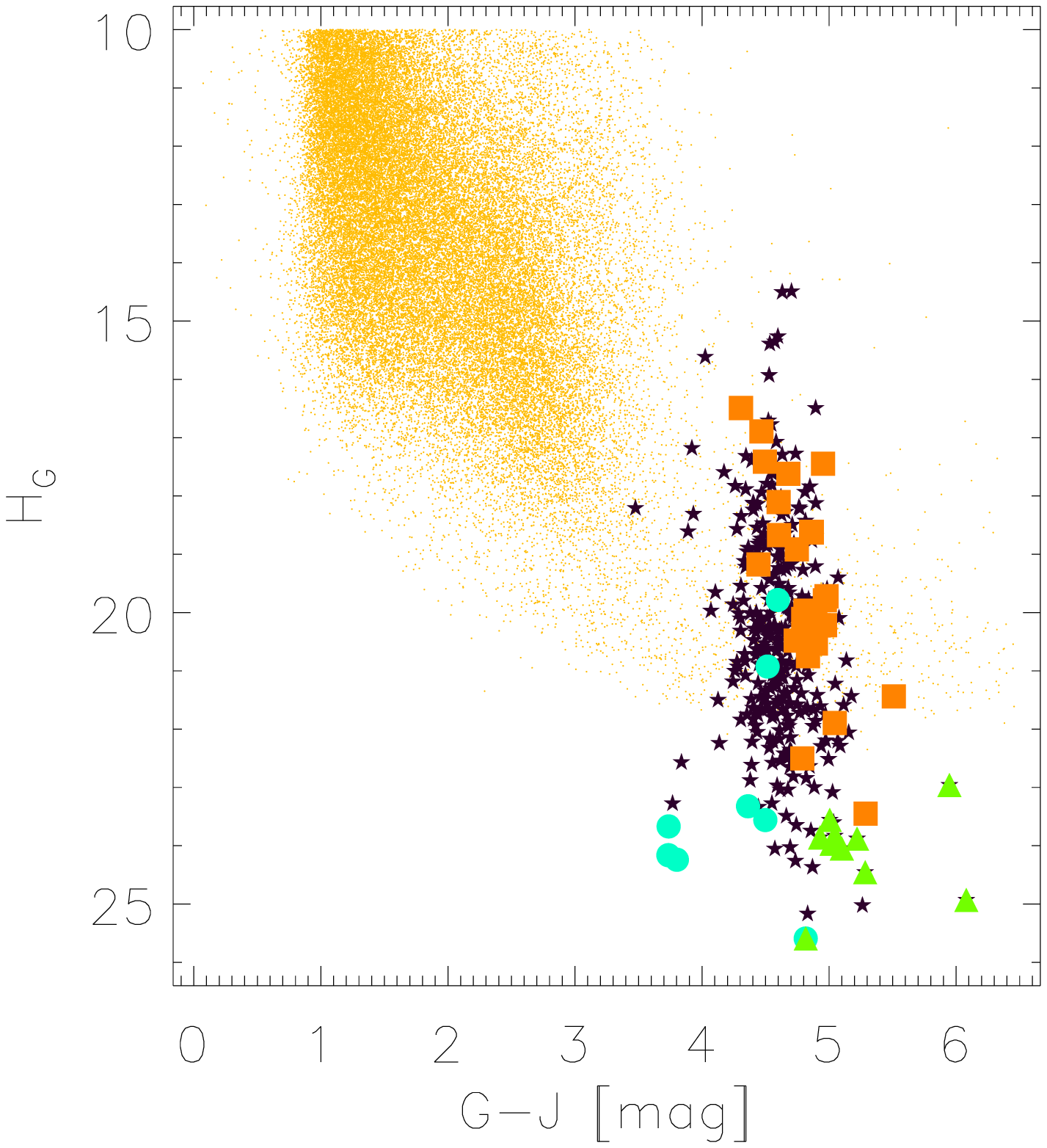}
\includegraphics[width=0.52\textwidth]{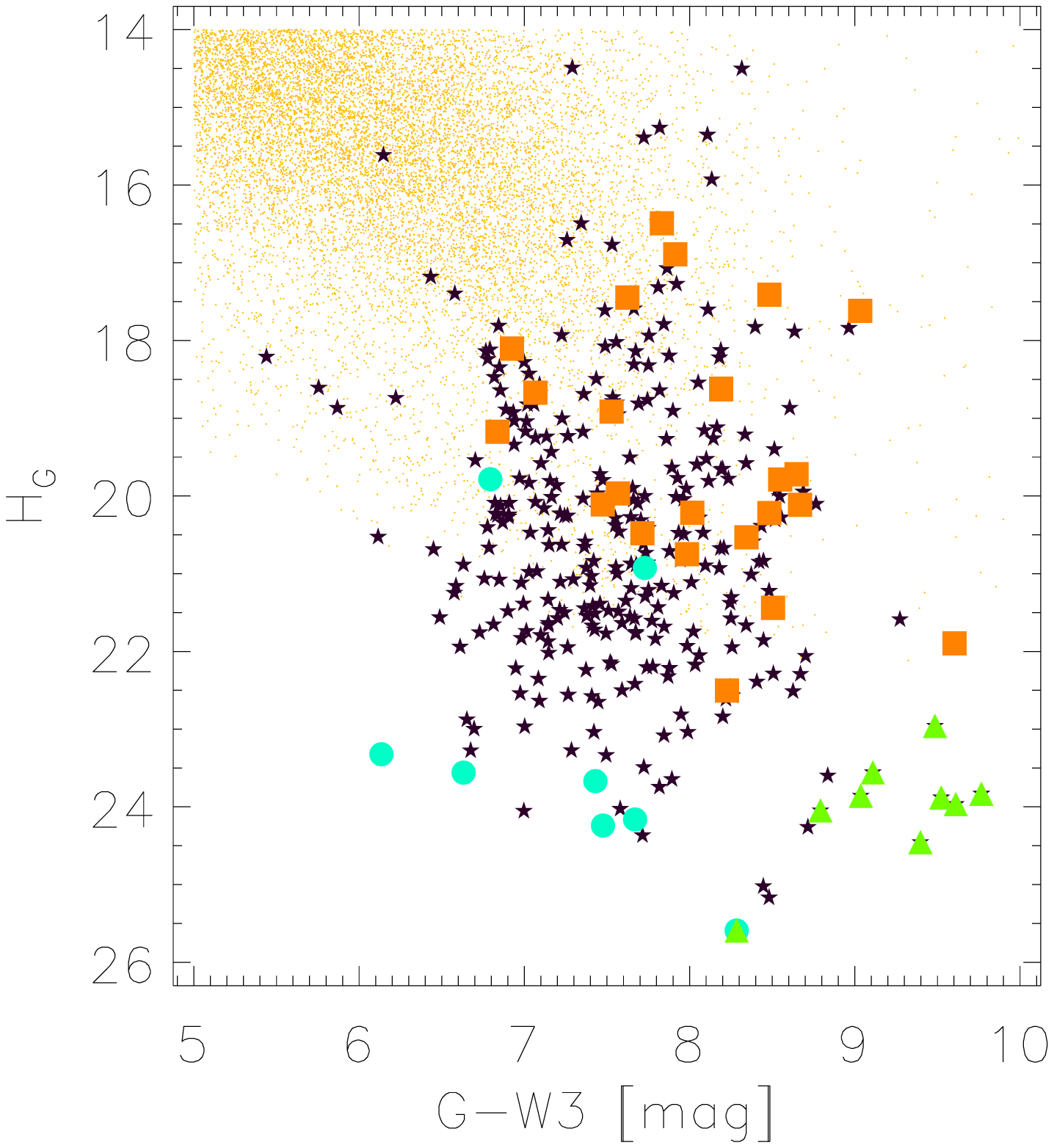}
\caption{Colour-reduced proper motion $H_G$ diagrams for $G-J$ and
  $G-W3$. Symbols have the same meaning as Fig.  \ref{COL1}. The top
  panel includes a significant part of the background objects, the
  bottom panel is a zoom on the region with LT dwarfs.  Additional
  colour-spectral type diagrams are found in Appendix~A.  }
\label{RPM1}
\end{figure}
\subsection{Reduced proper motion diagrams}

%

The reduced proper motion in the $G$ magnitude is defined as
\begin{equation} H_G = G + 5 +5 \log{\mu},
\end{equation}  where $\mu$ is the total
proper motion \citep{1972ApJ...173..671J}.
The $H_G$ is correlated with the absolute magnitude in $G$ via the tangential
velocity. Since objects in the solar neighbourhood tend to share the same
rotational velocity around the Galactic centre as the Sun, their kinematics
are restricted to velocity ellipsoids. Also since spectral types are also
correlated to absolute magnitudes, a plot of $H_G$ versus a surrogate of
spectral type provides a reduced proper motion diagram, which is a powerful
tool for isolating different stellar populations 
\citep[e.g. ][]{2009AJ....137....1F,2011AJ....142..138L,2012A&A...539A..86J}.

In Fig. \ref{RPM1} we plot the reduced proper motion $H_G$ vs. 
$G-J$ and $G-W3$ for the LT dwarfs identified in
the DR1, and include anonymous objects within 2\,arcmin of each LT
dwarf with proper motions calculated the same way (i.e. \G~DR1 -
2MASS).  The plots with other $G$ and 2MASS \& AllWISE colours are
included in the Appendix~A.
The anonymous objects trace
out the locus of the combined thin and thick disks, while the majority
of LT dwarfs clump in a relatively unoccupied region.  The subdwarfs
tend to occupy the lower part of the LT cloud while the young objects
are on the right hand edge.

There are 12 high proper motion objects with $H_G > 24.0$ mag listed in Table
\ref{largeH}, these include subdwarfs and objects later than L6.  For the
early L normal dwarfs a high $H_G$ would be an indication of a subdwarf
nature, however, as $H_G$ is a compound variable including both intrinsic
magnitude and kinematical properties it may be that it has a high velocity
because of encounters that have led to an unusually high value for $H_G$. The
distance and other parameters from \G~DR2 will clarify the nature of these
objects.

\begin{table}
  \begin{center}
    \caption[]{Objects with $H_G > 24.0$. }
    \label{largeH}
    \begin{tabular}{lllc}
      \hline
      \hline
      \noalign{\smallskip}
      Long name   & Opt SpT    & NIR SpT    &  $H_G$ [mag]          \\
      \noalign{\smallskip}
      \hline
      \noalign{\smallskip}
      ULAS J033350.84+001406.1  &  L0 sd$^1$ &   L0 sd$^2$ &  24.2\\ 
      DENIS J081730.0-615520    &    ...     &      T6$^3$ &  24.9\\ 
      2MASS 11555389+0559577    &    ...     &    L7.5$^4$ &  24.0\\ 
      2MASS 11582077+0435014    &  L7 sd$^5$ &  L7 sd$^5$  &  25.6\\ 
      SDSSp J120358.19+001550.3 &     L3$^6$ &      L5$^7$ &  24.4\\ 
      2MASS 12074717+0244249    &     L8$^8$ &      T0$^9$ &  24.5\\ 
      2MASS 12304562+2827583    &    ...     &   L1$^{10}$ &  24.3\\ 
      ULAS J124425.90+102441.9  &    ...     & L0.5 sd$^2$ &  24.2\\ 
      2MASSW J1411175+393636    & L1.5$^{11}$ &   L1.5$^7$ & 24.1\\ 
      2MASSW J1515008+484742    &   L6$^{12}$ &   L6$^{13}$ &  25.2\\ 
      2MASS 22114470+6856262    &        ... &     L2$^5$ &  24.0\\ 
      2MASS 22490917+3205489    &   L5$^{12}$ &       ...  &  25.0\\ 
      \noalign{\smallskip} 
      \hline
    \end{tabular}
  \end{center}
Spectral type references:   $^1$\cite{2010MNRAS.404.1817Z},
  $^2$\cite{2012A&A...542A.105L},
  $^3$\cite{2010APJ...718L..38A},
  $^4$\cite{2004AJ....127.3553K},
  $^5$\cite{2010APJS..190..100K},
  $^6$\cite{2000AJ....119..928F},
  $^7$\cite{2014APJ...794..143B},
  $^8$\cite{2002AJ....123.3409H},
  $^9$\cite{2006APJ...639.1095B},
  $^{10}$\cite{2009AJ....137..304S},
  $^{11}$\cite{2000AJ....120..447K},
  $^{12}$\cite{2007AJ....133..439C},
  $^{13}$\cite{2003IAUS..211..197W}.
\end{table}


\subsection{Hertzsprung-Russell diagrams}

\begin{figure*}
\centering
\includegraphics[width=0.49\textwidth]{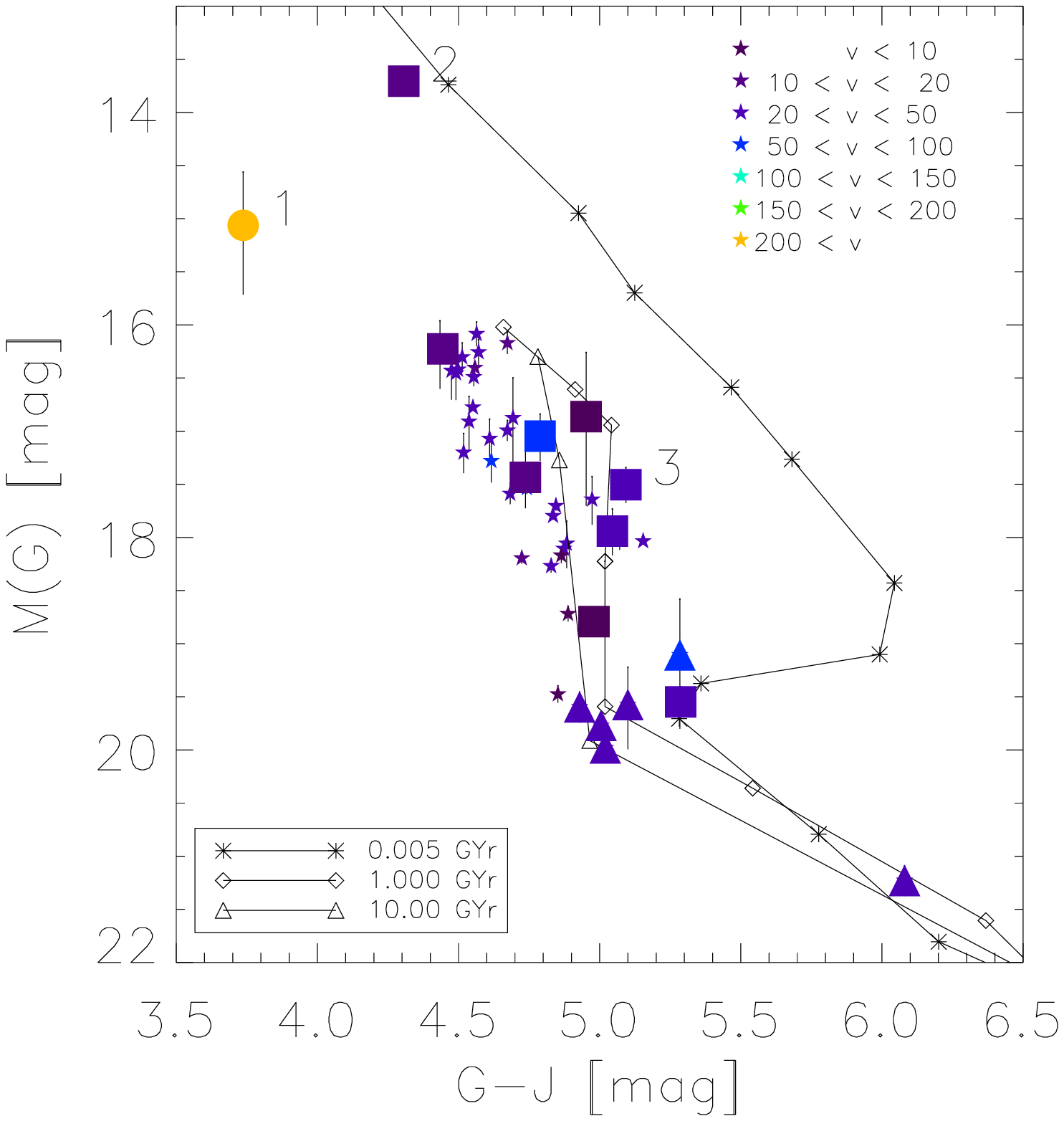}
\includegraphics[width=0.49\textwidth]{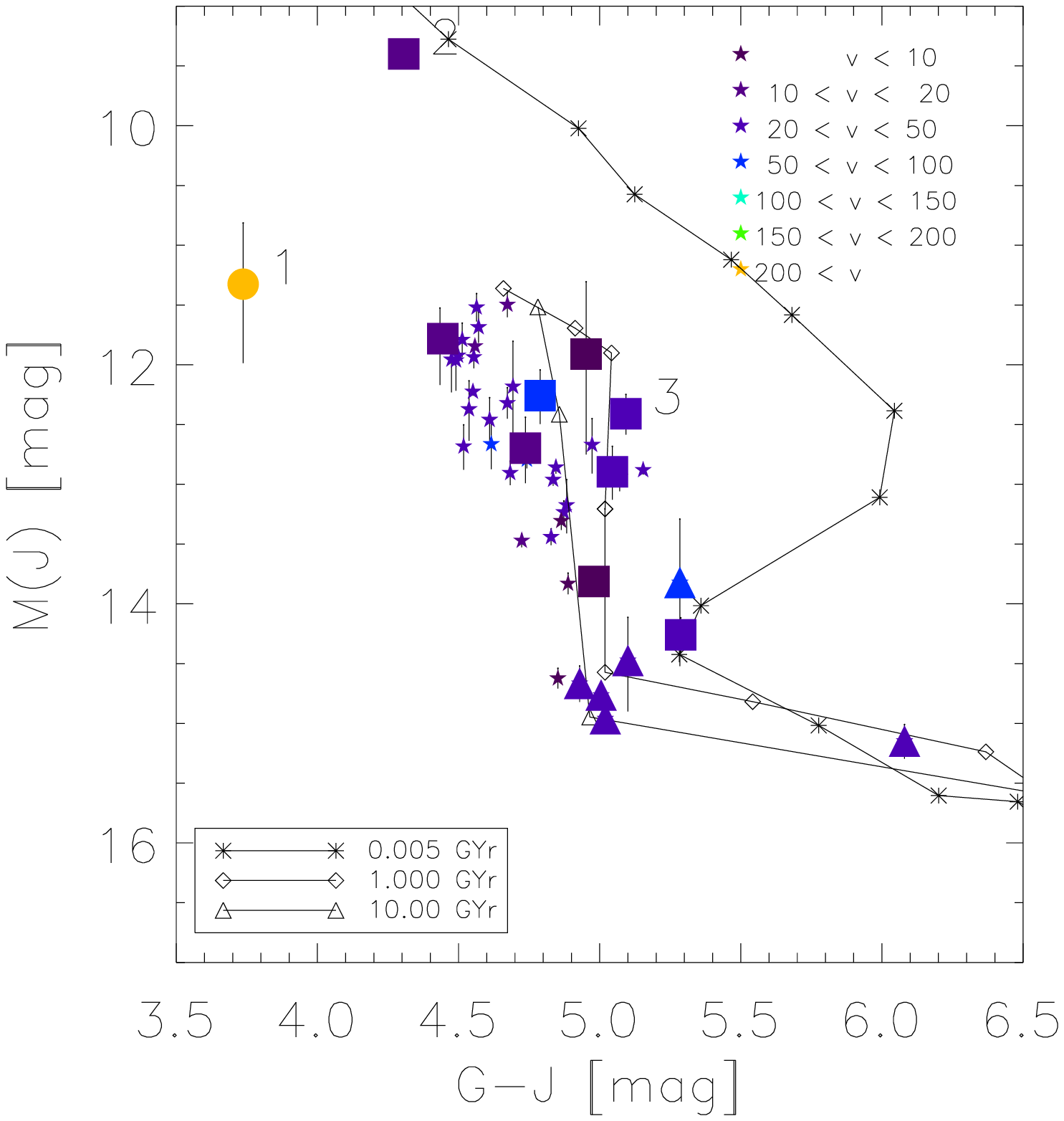}
\caption{Absolute magnitude in $G$ and $J$ bands vs. $G-J$. The colour of
  symbols indicate the velocity as given in the legend. Filled circles are
  subdwarfs, the squares are noted as young and triangles are objects later
  than L7, small stars are other L0--6 dwarfs. The objects labelled 1,2,3 are
  J1256-0224, TWA 27 A and Kelu-1 A respectively.}
\label{HRJ}
\end{figure*}

\begin{figure*}
\centering
\includegraphics[width=0.49\textwidth]{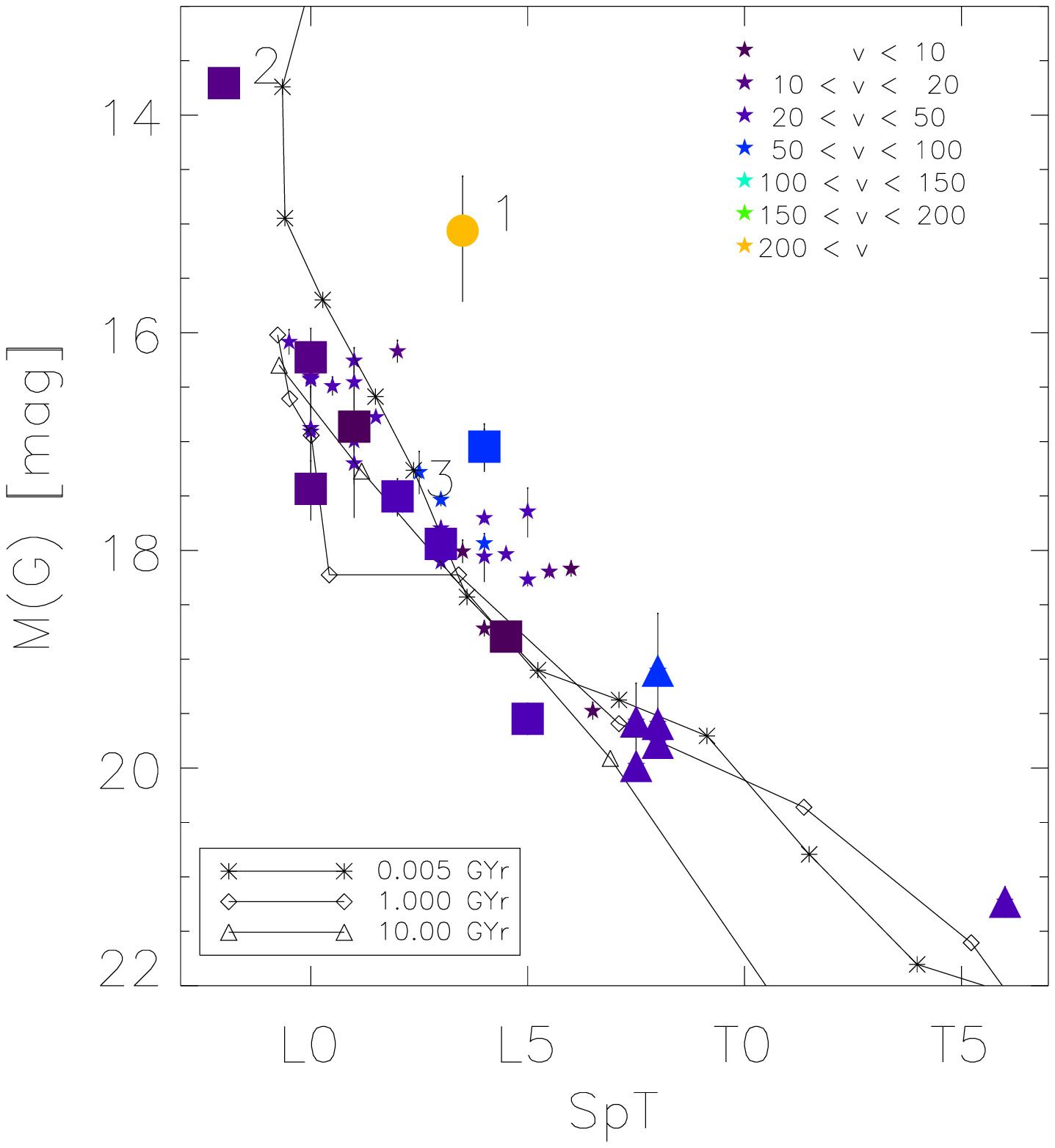}
\includegraphics[width=0.49\textwidth]{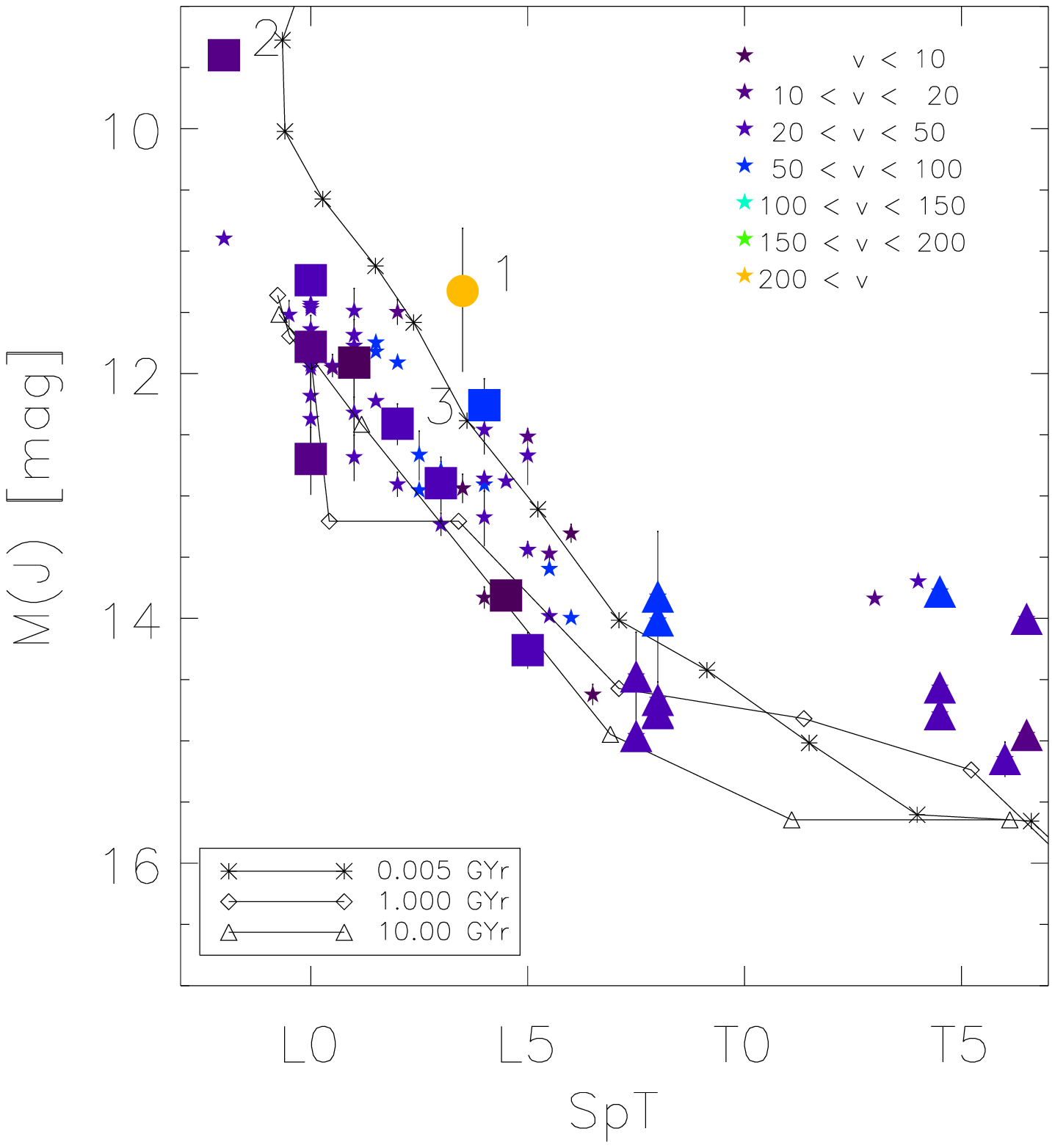}
\caption{Absolute magnitude in $G$ and $J$ bands vs. spectral types. The colour
  and shape of the  symbols are as given in Fig. \ref{HRJ}. The increased
  number of objects in the right panel are the LT dwarfs in CPM systems assuming
  the TGAS parallax which is only possible in this panel as we do not have $G$
  band magnitudes for the CPM LT objects. }
\label{HRSP}
\end{figure*}

For the \NHRD~targets that have published parallaxes, we can determine
absolute magnitudes and tangential velocities. In Fig. \ref{HRJ} and
\ref{HRSP} we plot the H-R diagrams with absolute $G$ magnitudes and both the
$G-J$ colour and the spectral type as surrogates for temperature. We have
colour coded the symbols to indicate the tangential velocities. There is a
spread of 0.7\,mag in absolute magnitude that, based on the propagation of the
formal errors, appears to be largely intrinsic. On each H-R diagram we have
included the most recent PHOENIX
isochrones\footnote{\url{https://phoenix.ens-lyon.fr/Grids/BT-Settl/CIFIST2011_2015/}}
\citep[][ and references therein]{2013MmSAI..84.1053A} for a 0.005, 1 and 10
Gyr, as shown in the legends.

The objects labelled 1,2,3 are 2MASS 12563716-0224522, TWA 27 A and Kelu-1
A. 2MASS 12563716-0224522 is a subdwarf \citep{2009A&A...494..949S} and TWA 27
A is a member of the young TW Hya association \citep{2002ApJ...575..484G}
hence the outlier positions. Kelu-1 is triple system, with Kelu-1 A a
spectroscopically identified L0.5+T7.5 \citep{2008arXiv0811.0556S} and Kelu-1
B a L3pec dwarf 300\,mas towards the southeast \citep{2005ApJ...634..616L} of
the primary double. \G~should resolve and detect both the double primary
system and the secondary so we assume that the object matched is the brighter
Kelu-1 A. Indeed if we use the $J$ magnitude of the combined system the object
takes an outlier position in the absolute magnitude as a function of $G-J$ but
not in absolute magnitude as a function of spectral type.


To Fig. \ref{HRSP} we have added the 28 CPM objects found in Section
\ref{section:cpm} with $J$ band magnitudes and adopting the parallax of the
TGAS CPM companion.  In some cases using the TGAS parallax is inappropriate,
e.g. in systems that are unbound, but the consistency in the H-R diagram
indicates this is not a bad assumption.  The fact that we can see
differences, even in this small sample with heterogeneous parallaxes, is a
taste of what to expect with the full GUCDS. The \G~DR2 is expected to furnish
precise parallaxes for most of the GUCDS and we will use the small differences
in colour and absolute magnitude trends to find more direct indicators of age,
metallicity and other physical properties.

%

\section{Conclusions}
\label{section:conclusion}
We have produced a catalogue of all known LT dwarfs with estimates of their $G$
magnitude and proper motions. We investigated \NCAT~known LT dwarfs identified
in the \G~DR1. The number of LT dwarfs identified is in line with the
expectation that \G~will directly observe around 1000 LT dwarfs. The addition
of new homogeneous optical magnitudes opens new possibilities for
interpretation, and the addition of distances and proper motions directly from
\G~will allow hypothesis testing on a statistically significant sample
size. In particular as we will be able to construct, for the first time,
volume limited complete samples.

The energy of LT dwarfs is primarily emitted in the near infrared bands, and
the energy being emitted in the $G$ band is from the very red edge of the
filter or is non-thermal. An example of non-thermal emission can be seen in
the 1997 flare of 2MASSW~J0149090+295613 which appears to have been for a
short period brighter in the optical than it is in the infrared in its
quiescent state \citep{1999ApJ...519..345L}.
\G~in its normal operation will make an average of more than 80 observations
per target with nine precise measures of $G$ during each observation. This will
be a well defined, well sampled dataset that will be able to constraint and
characterise the occurrences of flares in LT dwarfs in the optical regime.

Even though many of these objects will be close to the detection limit of \G~
their relative closeness and nominal \G~precision will allow us to calculate
tangential velocities with high precisions of meters-per-second. This
precision will in turn allow us to co-locate them with local moving groups,
streams and common proper motion systems that will provide a wealth of
constraints on the physical properties of LT dwarfs. This is evidenced by the
diverse locations of young and subdwarf LT objects. 

We have found \NEWCPMS~candidate CPM systems by a comparison of our input
catalogue to the \G~TGAS subset. The ability to identify CPM pairs will allow
us to push down towards the coolest brown dwarfs and the \G~results will be
crucial to fully characterise the systems and constrain objects that will be
too faint for \G. Eventually these benchmark GUCDS objects with age,
metalicity and distance constraints provided by the brighter companion or by
the parent association will be the sample to constrain our global picture of
UCDs.  Ultimately, we hope the GUCDS will allow us to identify observational
spectral and colour indicators for the direct determination of physical
properties like age and mass.


\section*{Acknowledgments}
RLS's research was supported by a visiting professorship from the Leverhulme
Trust (VP1-2015-063).  FM/HRAJ/DJP acknowledge support from the UK's Science and Technology
Facilities Council grant number ST/M001008/1. DB acknowledges support from the
Spanish grant ESP2015-65712-C5-1-R. JAC acknowledges support from 
Spanish grant AYA2015-74551-JIN. The early work on this project was
supported by the Marie Curie 7th European Community Framework Programme grant
n.247593 {\it Interpretation and Parameterization of Extremely Red COOL
  dwarfs} (IPERCOOL) International Research Staff Exchange Scheme.

This research has made use of:
observations collected at the European Organisation for Astronomical
Research in the Southern Hemisphere under ESO programme 097.C-0592(A);
data from the European Space Agency mission
 \G\footnote{\url{http://www.cosmos.esa.int/gaia}}, processed by the  \G~Data
Processing and Analysis
Consortium\footnote{\url{http://www.cosmos.esa.int/web/gaia/dpac/consortium}},
funded by national institutions participating in the  \G~Multilateral
Agreement and in particular the support of ASI under contract I/058/10/0 (\G~
Mission - The Italian Participation to DPAC);
the SIMBAD database and the VizieR catalogue access tool, provided by CDS,
Strasbourg, France;
data products from the {\it Wide-field Infrared Survey Explorer}, which is a
joint project of the University of California, Los Angeles, and the Jet
Propulsion Laboratory/California Institute of Technology, and NEOWISE, which
is a project of the Jet Propulsion Laboratory/California Institute of
Technology; WISE and NEOWISE are funded by the National Aeronautics and Space
Administration;
the Two Micron All Sky Survey, which is a joint project of the University of
Massachusetts and the Infrared Processing and Analysis Center/California
Institute of Technology, funded by the National Aeronautics and Space
Administration and the National Science Foundation;
the SpeX Prism Spectral Libraries, maintained by Adam
Burgasser\footnote{\url{http://pono.ucsd.edu/~adam/browndwarfs/spexprism}};
and the M, L, T and Y dwarf compendium maintained by Chris Gelino, Davy
Kirkpatrick and Adam Burgasser.
%


\bibliographystyle{mnras}
\bibliography{refs} 

\end{document}